%% file: seq.tex
\documentclass[10pt,journal,compsoc]{IEEEtran}
\usepackage[none]{hyphenat}
\usepackage{color,graphicx}
\usepackage{url}
\usepackage{amssymb,amsmath,dsfont}
\usepackage{amsmath,bm}
\usepackage{subfigure}
\usepackage{multirow}
\usepackage{epsfig}
\usepackage{dblfloatfix}
\usepackage[vlined,ruled,linesnumbered]{algorithm2e}
\usepackage{subfigure,multirow}
\usepackage{amsmath}
\usepackage{epsfig}
\usepackage{ctable}
\usepackage{threeparttable}
\usepackage{enumerate}
\usepackage{amsmath}
\usepackage{amsfonts}
\usepackage{color}
\usepackage{multirow}
\usepackage{graphicx}
\usepackage{balance}
\usepackage{paralist}
\usepackage{CJKutf8}
\usepackage{url}
\usepackage{amssymb}
\usepackage{multirow}

\emergencystretch=\maxdimen
\newcommand{\nop}[1]{}

\newtheorem{problem}{Problem~}

\newtheorem{definition1}{Definition}

\usepackage{balance}

\def\BibTeX{{\rm B\kern-.05em{\sc i\kern-.025em b}\kern-.08emT\kern-.1667em\lower.7ex\hbox{E}\kern-.125emX}}

\begin{document}

\title{Outdoor Position Recovery from Heterogeneous Telco Cellular Data}


\author{Yige Zhang, Weixiong Rao, Kun Zhang and Lei Chen
\IEEEcompsocitemizethanks{
\IEEEcompsocthanksitem Yige Zhang and Weixiong Rao are with School of Software Engineering, Tongji University, Shanghai, China. \protect
E-mail: \{wxrao, 1610832\}@tongji.edu.cn\}

\IEEEcompsocthanksitem Kun Zhang is with Department of Philosophy
Carnegie Mellon University, Pittsburgh, United States. \protect
E-mail: kunz1@cmu.edu
\IEEEcompsocthanksitem Lei Chen is withDepartment of Computer Science and Engineering, Hong Kong University of Science and Technology, Kowloon, Hong Kong, China. \protect
E-mail: leichen@cse.ust.hk
}
\thanks{}}

\maketitle
\begin{abstract}
Recent years have witnessed unprecedented amounts of data generated by telecommunication (Telco) cellular networks. For example, measurement records (MRs) are generated to report the connection states between mobile devices and Telco networks, e.g., received signal strength. MR data have been widely used to localize outdoor mobile devices for human mobility analysis, urban planning, and traffic forecasting. Existing works using first-order sequence models such as the Hidden Markov Model (HMM) attempt to capture spatio-temporal locality in underlying mobility patterns for lower localization errors. The HMM approaches typically assume stable mobility patterns of the underlying mobile devices. Yet real MR datasets exhibit heterogeneous mobility patterns due to mixed transportation modes of the underlying mobile devices and uneven distribution of the positions associated with MR samples. Thus, the existing solutions cannot handle these heterogeneous mobility patterns. we propose a multi-task learning-based deep neural network (DNN) framework, namely \textsf{PRNet}$^+$, to incorporate outdoor position recovery and transportation mode detection. To make sure the framework work, \textsf{PRNet}$^+$ develops a feature extraction module to precisely learn local-, short- and long-term spatio-temporal locality from heterogeneous MR samples. Extensive evaluation on eight datasets collected at three representative areas in Shanghai indicates that \textsf{PRNet}${^{+}}$ greatly outperforms state-of-the-arts.
\end{abstract}
\input{01-introduction}
\input{02-pre}

\input{03-method}

\input{04-eva}
\input{06-con}

\bibliographystyle{abbrv}
\bibliography{deep}

\scriptsize
\begin{IEEEbiography}[{\includegraphics[width=0.8in,height=1.0in,clip,
keepaspectratio ]{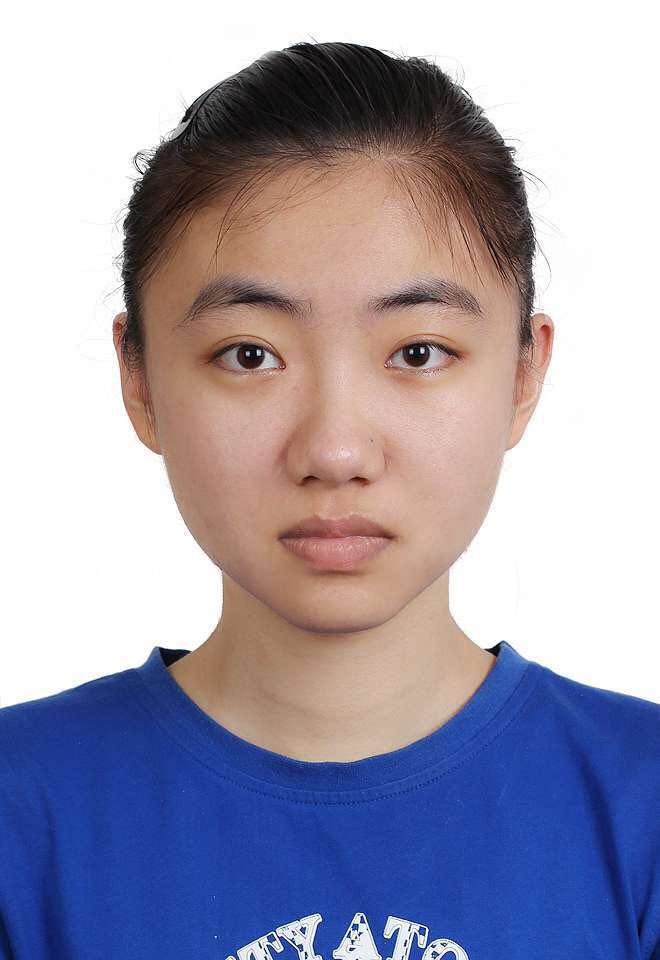}}]
{Yige Zhang} is a PhD student in School of Software Engineering, Tongji University, China since Sept 2016, and received the B.Sc degree of Software Engineering from Tongji University in July 2016. Her research interests focus on mobile computing and machine learning.
\end{IEEEbiography}\vspace{-4ex}

\begin{IEEEbiography}[{\includegraphics[width=0.8in,height=1.0in,clip,
keepaspectratio ]{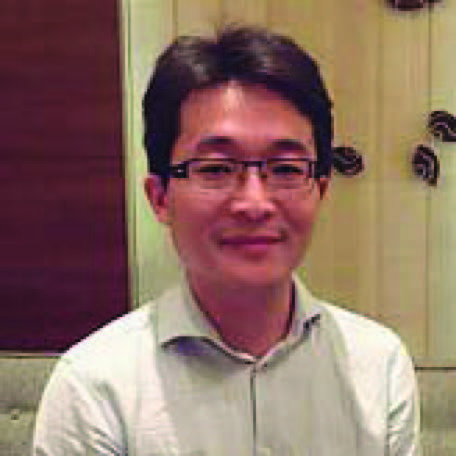}}]
{Weixiong Rao} received his Ph.D degree from The Chinese University of Hong Kong in 2009. After that, he worked for Hong Kong University of Science and Technology (2010), University of Helsinki (2011-2012), and University of Cambridge Computer Laboratory Systems Research Group (2013) as Post-Doctor researchers. He is a full Professor in School of Software Engineering, Tongji University, China (since 2014). His research interests include mobile computing and spatiotemporal data science, and is a member of CCF, ACM  and IEEE.
\end{IEEEbiography}
\begin{IEEEbiography}[{\includegraphics[width=0.8in,height=1.0in,clip,
keepaspectratio ]{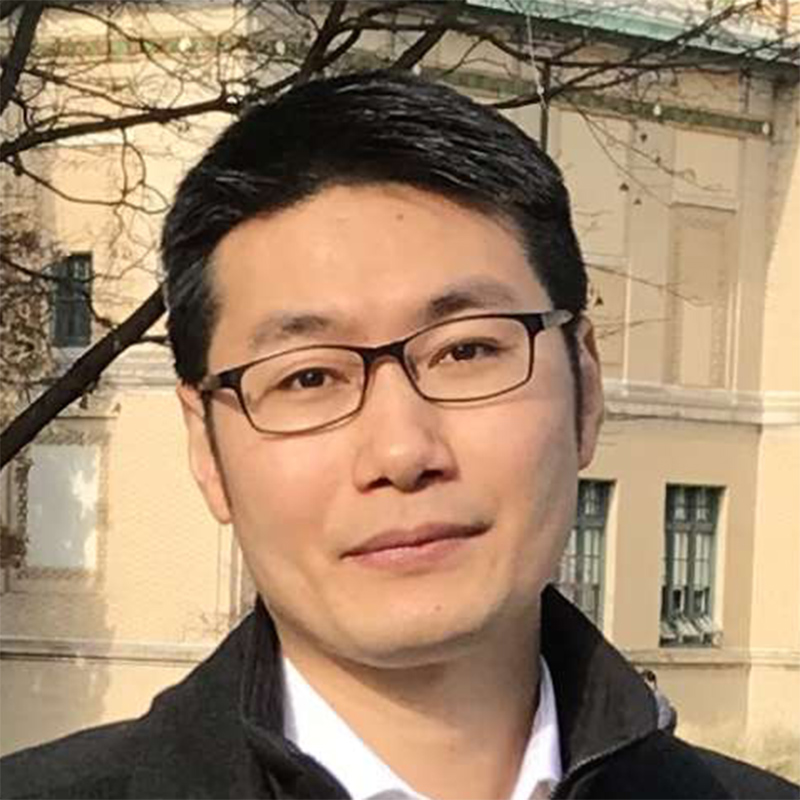}}]
{Kun Zhang} is an associate professor in the philosophy department and an affiliate faculty member in the machine learning department at Carnegie Mellon University. He received his Ph.D. degree from The Chinese University of Hong Kong in 2005. His research interests lie in machine learning and artificial intelligence, especially in causal discovery, causality-based learning, and general-purpose artificial intelligence. He coauthored a best student paper at UAI 2010, received the best benchmark award of the causality challenge 2008, and coauthored a finalist best paper at CVPR 2019. He has served as an area chair or senior program committee member for major conferences in machine learning or artificial intelligence, including NeurIPS, UAI, ICML, AISTATS, AAAI, and IJCAI, and has organized various academic activities to foster interdisciplinary research in causality. 
\end{IEEEbiography}

\begin{IEEEbiography}[{\includegraphics[width=0.8in,height=1.0in,clip, keepaspectratio ]{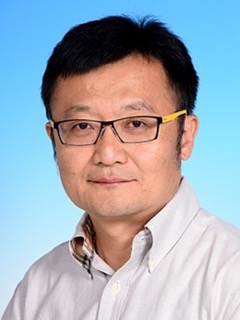}}]
{Lei Chen} is a full professor in the Department
of Computer Science and Engineering, Hong Kong University of Science and Technology. His research interests include crowdsourcing over social media, social media analysis,
probabilistic and uncertain databases, and privacy-preserved data publishing. The system
developed by his team won the excellent
demonstration award in VLDB 2014. He got
the SIGMOD Test-of-Time Award in 2015. He
is a member of the VLDB endowment and
ACM Distinguished Scientist.
\end{IEEEbiography}\vspace{-4ex}

\end{document}

%% file: 01-introduction.tex
\section{Introduction}
Recent years have witnessed unprecedented amounts of data generated by telecommunication (Telco) cellular networks. For example, when mobile devices make phone calls or access data services, measurement records (MRs) are generated to report connection states, e.g., received signal strength of mobile device during each call/session. In a modern urban city, Telco data generated by Telco radio and core equipment account to 2.2TB per day \cite{DBLP:conf/mdm/CostaZ18}. Massive Telco historical data (e.g., MRs) have been widely used to understand human mobility \cite{BeckerCHILMRUVV13,zhao2015explaining} and facilitate  applications such as urban planning and traffic forecasting \cite{BeckerCHLUVV11}, due to the unique advantages -- Telco data can be collected cheaply, frequently, and on a global scale.

The key to model human mobility and the applications above is to capture outdoor positions of mobile users. The position recovery problem, i.e., precisely localizing outdoor mobile devices from Telco historical data (e.g., MRs), has attracted intensive research interests. A simple approach adopted by Google MyLocation \footnote{\url{http://www.google.com/mobile/maps}} is to approximate outdoor locations by the positions of cellular towers connected to mobile devices. This method suffers from median errors of hundreds and even thousands of meters. More recently, by using the sparse geo-tagged MR samples as training data, machine learning-based Telco localization approaches such as Random Forest \cite{ZhuLYZZGDRZ16} and Hidden Markov Model (HMM) \cite{DBLP:conf/icc/IbrahimY11,DBLP:conf/conext/LeontiadisLKSWP14,RayDM16,8939387} are exploited to recover outdoor locations of mobile devices with the median errors of tens of meters. Such HMM approaches usually divide the area of interest into grid cells and treat the grid cells as hidden states and MR signal features as observations. When people holding mobile devices are moving around, these HMM-based works \cite{DBLP:conf/icc/IbrahimY11,DBLP:conf/conext/LeontiadisLKSWP14, RayDM16,8939387} capture spatio-temporal locality and frequently lead to better result than single point-based approaches \cite{ZhuLYZZGDRZ16} which independently localize individual MR samples.

However, the HMM approaches above do not work well if we cannot  precisely capture the distribution of emission and transition probabilities on real MR data. It is particularly true because MR data, if not geo-tagged, indicate rather coarse location information and frequently exhibit heterogeneous mobility patterns (e.g., caused by uneven distributions of the positions of MR samples and mixed transportation modes of underlying mobile devices). For example, due to the various timestamp intervals between neighbouring MR samples, the spatial distance between two neighbouring MR samples with a time interval of one minute could be much smaller than the one of ten minutes. Even with the same time interval (e.g., one minute), the distance between two neighbouring MR samples in the driving mode is significantly greater than the one in the walking mode. {If we still take the HMM approaches above for illustration, the transition probabilities across hidden states (i.e., grid cells) do not take into account the mixed transportation modes and various timestamp intervals of MR samples, resulting in the poor localization performance.} Thus, these HMM approaches cannot precisely capture spatio-temporal locality and there still exist significant potentials for better accuracy.

In addition, understanding human transportation modes is helpful for many applications such as urban planning, transportation management and precision advertisement. Many existing methods \cite{DBLP:journals/corr/abs-1804-02386,DBLP:conf/ijcai/SongKS16,DBLP:conf/gis/StennethWYX11, DBLP:journals/tweb/ZhengCLXM10} exploit GPS trajectory data to infer users' transportation modes, and do not work well on Telco data (e.g., MR samples). In addition, Telco data usually cover much higher population and more areas, and it is meaningful to detect transportation modes on Telco data for urban sensing. On the other hand, the transportation modes of underlying mobile devices are helpful to capture moving speed and then estimate the moving distance of mobile devices within a certain time period. Thus, if the transportation modes are known, we have chance to improve outdoor position recovery.

{To this end, we propose a multi-task learning-based deep neural network (DNN) framework, namely \textsf{PRNet}$^+$, to incorporate the two tasks of outdoor position recovery and transportation mode detection. To make sure the framework work, \textsf{PRNet}$^+$ develops a feature extraction module to precisely learn common local-, short- and long-term spatio-temporal locality from heterogeneous MR samples, with help of the power of a convolutional neural network (\textsf{CNN}), a recently developed sequence learning model, i.e., long short-term memory cells (\textsf{LSTM}), and attention mechanism. The learned features are then used to solve the two tasks. Specifically, \textsf{PRNet}$^+$ 1) allows the various-length sequences of MR samples, such that the two components (CNN and LSTM) are able to capture spatial locality from the samples within each MR sequence, 2) exploits two attention mechanisms for the time-interval between neighbouring MR samples, together with the one between neighbouring MR sequences, to capture temporal locality, and 3) incorporates the detected transportation modes and predicted locations of heterogeneous MR data into a joint loss for better result. As a summary, we make the following contributions.}


\emph{1}) To the best of our knowledge, this is the first attempt to fully employ the power of multiple DNN models to automatically learn spatio-temporal features from heterogeneous MR samples. Compared to the existing work on HMM-based localization, \textsf{PRNet}$^+$ can train a spatio-temporal sequence prediction model from heterogeneous MR data for accurate location recovery.


\emph{2}) {\textsf{PRNet}$^+$ properly leverages a multi-task learning framework to solve the two tasks of outdoor position recovery and transportation mode detection. Since the two tasks share the same features learned from heterogeneous MR data via the developed feature extraction module, we have chance to avoid high overheads to learn features independently for each task. 


\emph{3}) We evaluate \textsf{PRNet}$^+$ against state-of-the-art counterparts by using eight datasets collected on three representative (core, urban and rural) areas in Shanghai, China. Our evaluation indicates that \textsf{PRNet}$^+$ significantly outperforms these counterparts.

The rest of this paper is organized as follows. Section \ref{sec:pre} first introduces preliminaries. Section \ref{sec:problem} then gives the problem setting and Sections \ref{sec:design}-\ref{s:prnet_training} present the design. Next, Section \ref{sec:eva} evaluates \textsf{PRNet}$^+$ and Section \ref{sec:related} reviews related works. Section \ref{sec:con} finally concludes the paper.

%% file: 02-pre.tex
\section{Preliminaries}\label{sec:pre}

\begin{table*}[hbt]
\scriptsize
\centering\caption{2G GSM MR Sample}\label{tab:mr}\vspace{-2ex}
\begin{tabular}{|ll|ll|ll|ll|ll|}
\hline
$\textbf{MRTime}$& 2018/4/23 9:20&	$\textbf{IMSI}$ &xxx&	$\textbf{SRNCID}$ &6188&	$\textbf{BestCellID}$ &26051&	$\textbf{LCS BIT}$&	300
\\\hline
$\textbf{RNCID\_1}$& 6188&	$\textbf{CellID\_1}$& 26051&$\textbf{AsuLevel\_1}$ &27&	$\textbf{SignalLevel\_1}$& 4&	$\textbf{RSSI\_1}$& -74.5 \\\hline
$\textbf{RNCID\_2}$&6188&	$\textbf{CellID\_2}$&27394&	$\textbf{AsuLevel\_2}$&10&	$\textbf{SignalLevel\_2}$&	3& $\textbf{RSSI\_2}$&-84.88\\\hline
$\textbf{RNCID\_3}$&6188&	$\textbf{CellID\_3}$&27377& $\textbf{AsuLevel\_3}$&	18&$\textbf{SignalLevel\_3}$&4&	$\textbf{RSSI\_3}$&-85.13 \\\hline
$\textbf{RNCID\_4}$&6188&	$\textbf{CellID\_4}$&27378&$\textbf{AsuLevel\_4}$&12&	$\textbf{SignalLevel\_4}$&4& $\textbf{RSSI\_4}$&-85.87\\\hline
$\textbf{RNCID\_5}$&	6182& $\textbf{CellID\_5}$&41139&	$\textbf{AsuLevel\_5}$&8&	$\textbf{SignalLevel\_5}$&3&	$\textbf{RSSI\_5}$&-88.88 \\\hline
$\textbf{RNCID\_6}$&6188&	$\textbf{CellID\_6}$&	27393&$\textbf{AsuLevel\_6}$&9&	$\textbf{SignalLevel\_6}$&3&	$\textbf{RSSI\_6}$&-90.22\\\hline
$\textbf{RNCID\_7}$&6182&	$\textbf{CellID\_7}$&44754&	$\textbf{AsuLevel\_7}$&9&	$\textbf{SignalLevel\_7}$&	3& $\textbf{RSSI\_7}$&-95 \\

\hline
\end{tabular}
\end{table*}

\textbf{Measurement Record (MR) Data}: MR data measure the connection states between mobile devices and neighbouring base stations. Table \ref{tab:mr} gives an example of 2G GSM MR collected by an Android phone. This sample contains a unique number (known as IMSI, International Mobile Subscriber Identity), connection time stamp (MRTime), up to 7 nearby base stations (RNCID and CellID), signal measurements such as AsuLevel, SignalLevel, and RSSI (radio signal strength indicator). The up to 7 base stations are frequently sorted by descending order in signal level and strength. Thus, the base station with the order index 1 (RNCID\_1 and CellID\_1) is with the strongest signal and typically selected as the primary serving base station to provide communication and data services for mobile devices. When a mobile device is moving out of the signal coverage range of a primary serving base station, the \emph{handoff} between base stations occurs to re-select a new primary serving base station for the mobile device.

\textbf{Telco Location Recovery}: Existing works on Telco position recovery are typically divided into three categories. Firstly, \emph{measurement-based methods} approaches localize mobile devices based on absolute point-to-point distance or angles \cite{Patwari2005Locating,swales1999locating,DBLP:conf/icassp/VaghefiGS11}. Triangulation localization, one of the widely used measurement-based approach, usually does not work well for 4G LTE MR data where frequently signal strengths of one or at most two cells are available. Yet triangulation localization requires signal strength regarding three and ideally four or more cells.

Secondly, \emph{fingerprinting-based methods} \cite{IbrahimY12,Caffery1998,MargoliesBBDJUV17} usually have better performance than measurement-based methods, by dividing an area of interest into small grids and representing each grid by an associated fingerprint \cite{Caffery1998}. NBL \cite{MargoliesBBDJUV17}, a recently proposed work, assumes that signal strengths of each neighbouring cell tower in the grid follow a Gaussian distribution. The online stage next adopts either Maximum Likelihood Estimation (MLE) or Weighted Average (WA) to localize mobile devices.

Lastly, some recent works \cite{DBLP:conf/gis/ShokryTY18,IbrahimY12, MargoliesBBDJUV17,ZhuLYZZGDRZ16,ZhangRX18} adopt \emph{machine learning (ML) techniques} such as Random Forest to build a localization model to maintain the correlations between the features extracted from MR samples and associated locations (e.g., GPS coordinates). The predicted locations could be either spatial regions (grid cells) or numeric GSP coordinates. ML techniques then train the corresponding multi-classifiers \cite{IbrahimY12, MargoliesBBDJUV17, DBLP:conf/gis/ShokryTY18} or regression models \cite{ZhuLYZZGDRZ16}. Recently, a deep neural network-based outdoor cellular localization system, namely DeepLoc \cite{DBLP:conf/gis/ShokryTY18}, has been proposed. DeepLoc mainly utilizes a data augmentor to handle data noise issue and to provide more training samples, and trains a deep learning model using the augmented samples for better localization result.

Note that ML techniques differ from measurement- and fingerprinting-based methods in terms of application scenarios. That is, ML techniques usually exhibit better results because ML techniques can leverage the rich information from various Telco data including (geo-tagged) MR data and configuration parameters (e.g., GPS coordinates) of base stations, Web log data, and etc. Instead, when measurement and fingerprinting-based methods work on {frontend} mobile devices for active localization, they may not exploit the rich data as ML-based localization and thus cannot achieve comparable accuracy.

%% file: 03-method.tex
\section{Problem Setting and Solution Overview}\label{sec:problem}
\subsection{Problem Setting}
Suppose that Telco operators have maintained a historical MR database. MR samples usually do not contain the accurate locations of mobile devices (identified by IMSI), and we expect to annotate MR samples with the accurate locations of these mobile devices. There are various ways to acquire the locations. For example, when mobile users are using vehicle navigation services and switching on GPS receivers in the mobile devices, the GPS coordinates of mobile devices are embedded in the URLs of mobile web logs. By extracting the GPS coordinates from such URLs, we establish the linkage between extracted GPS coordinates and MR samples in light of IMSI and timestamp. In this way, MR samples are tagged by the linked GPS coordinates. 

Nevertheless, mobile users frequently switch off GPS receivers on mobile devices e.g., for energy saving, and mobile Web logs contain rather sparse GPS coordinates. Thus, the majority of Telco MR data do not be tagged with the associated GPS coordinates. To tackle this issue, we first train a machine learning model by using the sparse geo-tagged MR samples as training data and next recover outdoor locations for non-tagged MR samples (testing data). Recall that single-point-based location recovery \cite{ZhuLYZZGDRZ16} may not capture spatio-temporal locality of underlying mobile devices. We thus expect to recover a trajectory of the locations predicted from a sequence of MR samples.

\begin{definition1}
\textbf{[MR Database $\mathcal{D}$]} A database $\mathcal{D}$ of MR samples is organized as follows. We first group the MR samples by IMSI and then sort the samples in each group by timestamp. In this way, each IMSI is with a series of sorted MR samples.
\end{definition1}

Suppose that every MR sample in $\mathcal{D}$ is with an associated GPS position and transportation mode. Then for each IMSI, we have a series of sorted MR samples and an associated trajectory of GPS positions. We are now ready to define the position recovery problem.

\begin{problem}\label{pb1}
\textbf{[Position Recovery Problem]} Given a database $\mathcal{D}$, by training a multi-task learning-based framework $\mathds{R}$, the position recovery problem aims to map a testing MR series to a GPS trajectory by $\mathds{R}$.
\end{problem}

\subsection{Challenges}
A baseline approach to address Problem \ref{pb1} is to exploit a sequence model such as the popular RNN or LSTM model. For example, we might divide every MR series into fixed-length windows of MR samples and then feed these windows of MR samples into the LSTM network for trajectory recovery. However, the baseline approach does not work well to capture spatio-temporal locality and suffers from high localization errors, due to the following challenges caused by heterogeneous MR samples.

1) \emph{Mixed transportation modes}: The simple assumption that the mobile device regarding a given MR series always moves by a certain transportation mode (e.g., walking) clearly does not hold. Depending on various transportation modes, the spatial distance between neighbouring MR samples differs significantly.


2) \emph{Irregular MR sampling rate}: Practically it is highly possible that the time intervals of some neighbouring MR samples within a real world MR series are very short (e.g., one minute) and yet those of other samples are rather long (e.g., ten minutes). Thus, assuming fixed time intervals between neighbouring MR samples is unreasonable.

3) \emph{Uneven density of deployed base stations}: Dense base stations are frequently deployed in urban areas and yet spare ones in rural areas. Consequently, the neighbouring MR samples within the MR series located in urban and rural areas, even with exactly the same spatial distance and time interval, could exhibit very different Telco signal handoff behavior between base stations.


\subsection{Solution Overview}
\begin{figure*}[th]
\begin{center}					
    \centerline{\includegraphics[width=16.0cm]{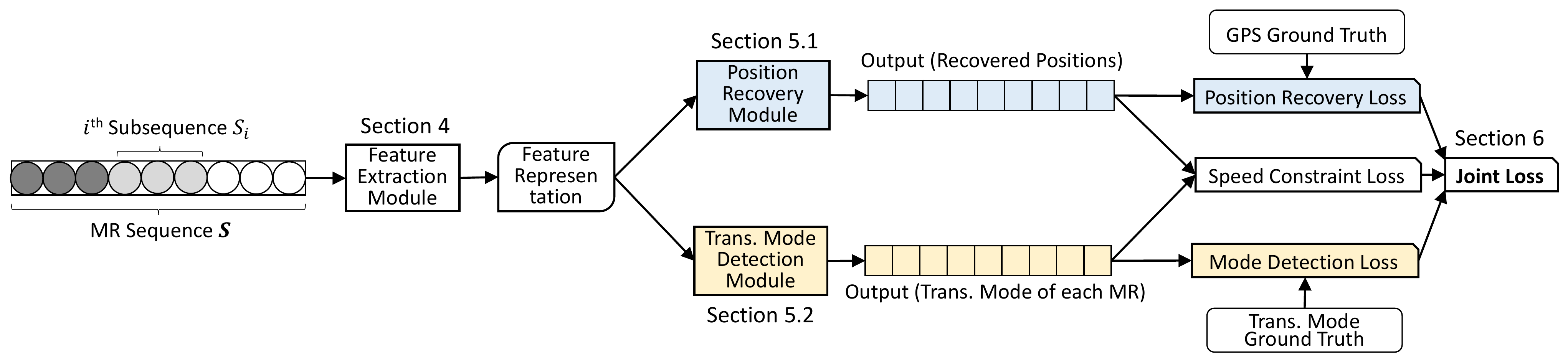}}\vspace{-2ex}
    \caption{The Architecture of \textsf{PRNet}$^{+}$} \label{fig:tprnet}
\end{center}\vspace{-4ex}
\end{figure*}


{To address above challenges, in Figure \ref{fig:tprnet}, we propose a multi-task learning-based solution framework, namely \textsf{PRNet}$^+$, which consists of three modules: \emph{1)} \textsf{Feature Extraction Module} to learn spatio-temporal features for an input MR sequence (we will soon give the formal definition of MR (sub)sequence), \emph{2)} \textsf{Position Recovery Module} to recover a trajectory of outdoor positions for an input MR sequence, and \emph{3)} \textsf{Mode Detection Module} to estimate the transportation mode for each MR sample within the MR sequence. Here, the two tasks of outdoor position recovery and transportation mode detection share the learned spatio-temporal features for better efficiency.}




{The key of \textsf{PRNet$^{+}$} is to train a multi-task learning framework, where the tasks of position recovery and transportation mode detection tasks share the same learned features. As shown in Figure \ref{fig:tprnet}, given the input of \textsf{PRNet}$^+$, i.e., one MR \emph{sequence} $S$, the feature extraction module learns spatio-temporal features, which are then fed into the two learning tasks (position recovery and mode detection) to generate final predictions, respectively. To train the multi-task learning framework, we design a join loss function consisting of three individual loss functions.} Thus, \textsf{PRNet}$^+$ differs from the previous work \textsf{PRNet} \cite{DBLP:conf/cikm/ZhangRZYZ19}, i.e., a single-task learning approach of outdoor position recovery. By incorporating both position recovery and mode detection tasks into the multi-task learning framework, \textsf{PRNet}$^+$ leads to much lower localization errors than \textsf{PRNet} alone. It makes sense because the learned transportation mode in \textsf{PRNet$^+$} directly indicates the moving speed constraint for better localization result.


\subsection{MR (Sub)sequence}\label{sec:mrseq}
It is not hard to find that the number of MR samples within a MR series varies across the associated mobile devices. To capture the spatial locality from the various-length MR series, we exploit the Telco signal \emph{handoff} between base stations as follows. Let us consider the Telco signal coverage range of a certain base station \emph{bs}, e.g., a circle with the radius around 1000 meters. The MR samples using \emph{bs} as the primary serving base station exhibit spatial locality because these samples are located within the circle. Thus, based on the stations in MR samples, we define the following MR (sub)sequences.

\begin{definition1}
\textbf{[MR Sequence $\mathcal{S}$]} Given a series of sorted MR samples regarding a certain IMSI, we divide the MR series into multiple sequences by primary serving base stations (\emph{RNCID\_1} and \emph{CellID\_1}). The MR samples in each sequence $\mathcal{S}$, sorted by timestamp, share the same primary serving base station and IMSI.
\end{definition1}

\begin{figure}[th]\vspace{-2ex}
\begin{center}					
    \centerline{\includegraphics[height=3.5cm]{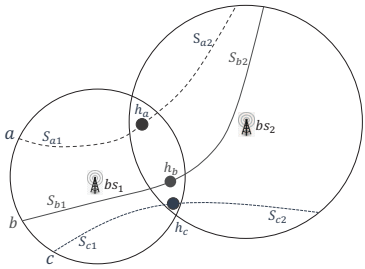}}\vspace{-2ex}
    \caption{Illustration of 6 MR Sequences} \label{fig:seq}\vspace{-2ex}
\end{center}\vspace{-2ex}
\end{figure}

Figure \ref{fig:seq} illustrates 6 MR sequences $S_{a1}, S_{a2},...,S_{c1}, S_{c2}$ generated from 3 MR series identified by the 3 mobile devices $a, b, c$ across the coverage range of two primary serving base stations \emph{bs}$_1$ and \emph{bs}$_2$. For a mobile device say $a$, the \emph{handoff point} $h_a$ divides its MR series into two sequences $S_{a1}$ and $S_{a2}$.

Beyond MR sequence, we define the following MR subsequences.
\begin{definition1}
\textbf{[MR Subsequence]} By the first non-serving base station (RNCID$\_2$ and CellID$\_2$), we divide a certain MR sequence $\mathcal{S}$ into multiple MR subsequences $S^{s}$. The MR samples in each $S$ share exactly the same non-serving base station (RNCID$\_2$ and CellID$\_2$), primary serving base station and IMSI.
\end{definition1}


Given the MR subsequences above, we expect that the MR samples within these subsequences exhibit spatio -temporal locality. In this way, \textsf{PRNet$^+$} has chance to capture the corresponding locality from these MR (sub)sequences for better localization and mode detection results. Note that the amount of MR samples within each MR sequence varies, depending upon the coverage radius of a base station, sampling rate, and transportation mode. In case a certain MR sequence contains very sparse samples, we could adaptively merge multiple neighbouring MR sequences into a single one which contains at least $\tau$ samples.


\section{Feature Extraction Module}\label{sec:design}
In this section, we first introduce the data model of feature extraction module in \textsf{PRNet$^+$}. For convenience, in the rest of this paper, we denote scalars by lowercase letters e.g., $a$, vectors by bold lowercase letters e.g., $\bm{a}$, matrices by bold upper-case letters e.g., $\bm{A}$, and tensors by bold upper-case letters e.g., $\mathbf{A}$. Specifically, we represent each MR sample by a matrix $\bm{X} \in \mathbb{R}^{F\times N}$, where $N$ is the number of base stations in this MR sample and $F$ is the number of MR features of each base station.  For the 2G MR sample in Table \ref{tab:mr}, we have $N=7$ stations (from the primary serving one to the 7-th station) and $F=7$ features (i.e., {RNCID}, CellID, latitude and longitude of the station identified by RNCID/CellID, ASULevel, SignalLevel, and RSSI).

We represent each MR sequence $\mathcal{S}$ by a sequence of MR feature matrices $\mathbf{X}=$ $\{\bm{X}_{1,1},...,\bm{X}_{i,j},...,\bm{X}_{q,|S_{q}}|\}$, where $\bm{X}_{i,j} \in \mathbb{R}^{F\times N}$ is the MR feature matrix of the $j$-th MR sample in the $i$-th subsequence $S_i$ with $1\leq i \leq q$ and $1\leq j\leq |S_q|$, where $q$ is the number of MR subsequences within $S$ and $|S_i|$ is the amount of MR samples in a MR subsequence $S_i$. Given the input MR sequence, we will learn a corresponding sequence of feature vectors $\bm{V}=\{\bm{v}_{1,1},...,$ $\bm{v}_{1,|S_{1}|},...,\bm{v}_{q,|S_{q}}|\}$.


\begin{figure}[th]
\begin{center}					
    \centerline{\includegraphics[height=5.2cm]{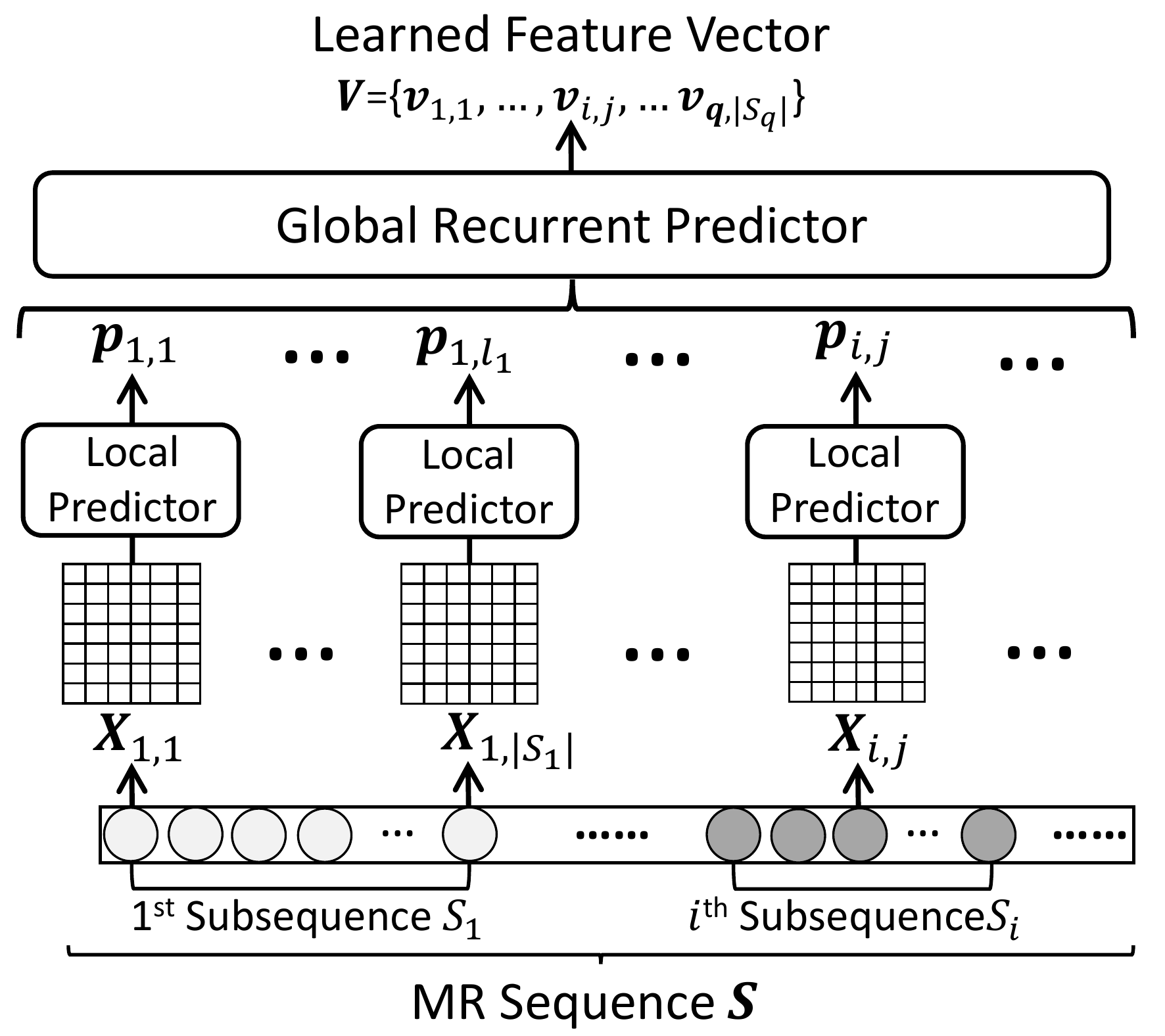}}
    \caption{Network Structure of \textsf{Feature Extraction Module}} \label{fig:overview}
\end{center}\vspace{-2ex}
\end{figure}


Figure \ref{fig:overview} gives the overview of \textsf{Feature Extraction Module}, consisting of two main components.

\textbf{\emph{1)} Local Single-Point Predictor} takes each individual matrix $\bm{X}_{i,j}$ as input and generates a corresponding hidden vector $\bm{p}_{i,j}$ as the output. The local predictor is composed of a convolution layer and a recurrent layer to capture the local dependencies from the MR features in $\bm{X}_{i,j}$.

\textbf{\emph{2)} Global Recurrent Predictor} takes a sequence of generated hidden vectors $\bm{P}=\{\bm{p}_{1,1},...,\bm{p}_{i,j},...,$ $\bm{p}_{q,|S_q|}\}$ as input and generates a sequence of shared feature representation vectors $\bm{V}$. This predictor consists of three layers: a bottom recurrent layer to learn short-term dependencies within each subsequence, an upper recurrent layer to capture the long-term dependencies among subsequences, and an output layer to mix the short- and long-term dependencies in order to generate the shared feature representation vectors of the input MR sequence.


\subsection{Local Predictor}
\begin{figure}[th]
\begin{center}					
    \centerline{\includegraphics[height=5cm]{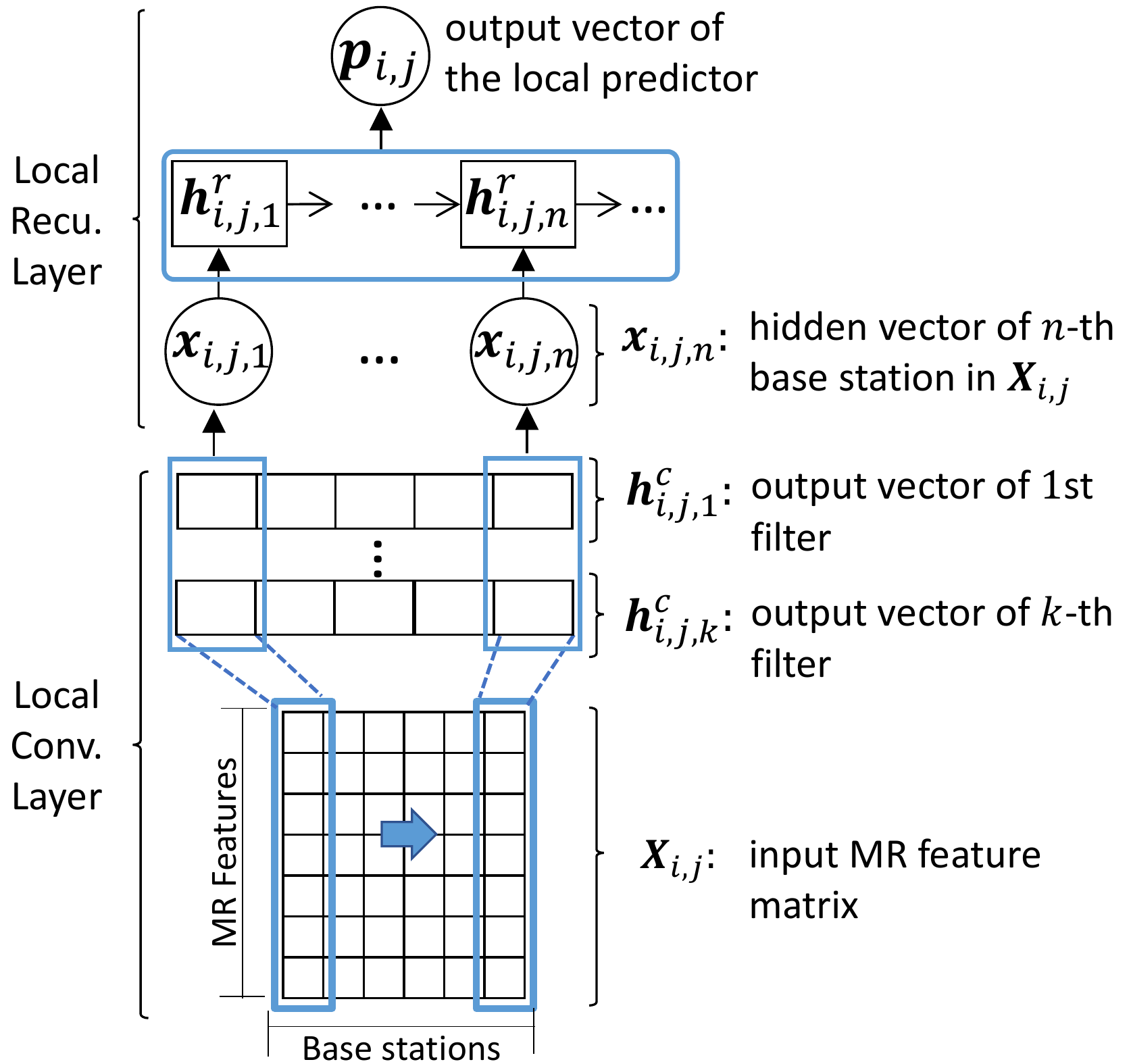}}\vspace{-2ex}
    \caption{Network Structure of Local Predictor} \label{fig:local_componet}
\end{center}\vspace{-3ex}
\end{figure}

Each MR sample contains the signal measurements of up to 7 base stations. Thus, in Figure \ref{fig:local_componet}, the convolution layer in the local predictor captures the local dependencies among the $F$ MR features from the input MR feature matrix $\bm{X}_{i,j}$ and generates an output feature vector $p_{i,j}$. Intuitively, the MR sample can be alternatively treated as a sequence of the $N$ base stations sorted by the associated signal measurements. Thus, the recurrent layer captures the local dependencies from the sequence.

\subsubsection{Local Convolution Layer}
Given an input $F \times N$ matrix $\bm{X}_{i,j}$, the convolution layer captures the local dependencies between MR features. Thus, the convolution layer adopts multiple one-dimensional filters with size $F \times 1$ where the height of each filter is equal to the number $F$. Since the size of the convolution filter is consistent with that of the feature vector of a certain base station in $\bm{X}_{i,j}$, the convolution operation can extract the local dependencies among multiple features of the base station. The output vector of the $k$-th filter throughout the input matrix $\bm{X}_{i,j}$ is:
\begin{equation}\small
\begin{aligned}
\bm{h}^{c}_{i,j,k}=ReLU(\bm{W}_{k}^{c} \circ \bm{X}_{i,j}+\bm{b}_{k}^{c}),
\end{aligned}
\end{equation}
where $\circ$ denotes a convolution operation and $ReLU(\cdot)$ is the activation function. The superscript $c$ indicates the \underline{c}onvolution layer. We set the stride of convolution operation to one and ensure that each output vector $\bm{h}^{c}_{i,j,k}$ has the size $N$.

Given $K$ filters, the output of the local convolution layer is a $K \times N$ matrix. For the $n$-th base station ($1\leq n\leq N$) within $\bm{X}_{i,j}$, the $K \times N$ output matrix has an associated row vector $\bm{x}_{i,j,n}\in \mathbb{R}^{K \times 1}$, treated as the latent feature vector of the $n$-th base station.

\subsubsection{Local Recurrent Layer} Given a $K \times N$ matrix above, the local recurrent layer treats it as a sequence of $N$ latent feature vectors. The intuition is that the order of these latent feature vectors indicates the inherent correlations among the $N$ base stations in an input MR sample. Thus, this local recurrent layer extracts the local dependencies from the latent feature vectors regarding $N$ base stations. We implement the local recurrent layer by LSTM \cite{DBLP:journals/neco/HochreiterS97} and compute the hidden state of the LSTM cells for the latent vector $\bm{x}_{i,j,n}$ as follows.

\begin{equation}\small
\begin{aligned}
\bm{z}_{i,j,n}^{r} &= \tanh (\bm{W}_{z}^{r}[\bm{h}_{i,j,n-1}^{r};\bm{x}_{i,j,n}]+\bm{b}_{z}^{r}), \\
\bm{f}_{i,j,n}^{r} &= \sigma(\bm{W}_{f}^{r}[\bm{h}_{i,j,n-1}^{r};\bm{x}_{i,j,n}]+\bm{b}_{f}^{r}), \\
\bm{g}_{i,j,n}^{r} &= \sigma(\bm{W}_{g}^{r}[\bm{h}_{i,j,n-1}^{r};\bm{x}_{i,j,n}]+\bm{b}_{g}^{r}), \\
\bm{c}_{i,j,n}^{r} &= \bm{f}_{i,j,n}^{r} * \bm{c}_{i,j,n-1}^{r}+\bm{g}_{i,j,n}^{r} * \bm{z}_{i,j,n}^{r}, \\
\bm{o}_{i,j,n}^{r} &= \sigma(\bm{W}_{o}^{r}[\bm{h}_{i,j,n-1}^{r};\bm{x}_{i,j,n}]+\bm{b}_{o}^{r}), \\
\bm{h}_{i,j,n}^{r} &= \bm{o}_{i,j,n}^{r} * \tanh \bm{c}_{i,j,n}^{r},
\end{aligned}
\end{equation}

where $*$ denotes element-wise multiplication and $\bm{h}_{i,j,n}$ is the hidden state of the $n$-th base station in $\bm{X}_{i,j}$. The superscript $r$ denotes a local \underline{r}ecurrent layer. The output of this layer, $\bm{p}_{i,j}\in\mathbb{R}^{(N \times d_l) \times 1}$, is the concatenation result of all hidden states, where $d_l$ denotes the number of hidden units in this local recurrent layer.

With the local predictor, \textsf{Feature Extraction Module} encodes each individual MR feature matrix $\bm{X}_{i,j}$ into a hidden feature vector $\bm{p}_{i,j}$. When an entire matrix sequence $\mathbf{X}$ is processed, the local predictor generates a sequence $\bm{P}=\{\bm{p}_{1,1},...,\bm{p}_{i,j},...,$ $\bm{p}_{q,|S_q|}\}$, which is fed into the global recurrent predictor to generate the sequence of shared feature representation vectors.

\subsection{Global Recurrent Predictor}
\begin{figure}[th]
\vspace{-1ex}
\begin{center}					
    \centerline{\includegraphics[height=5cm]{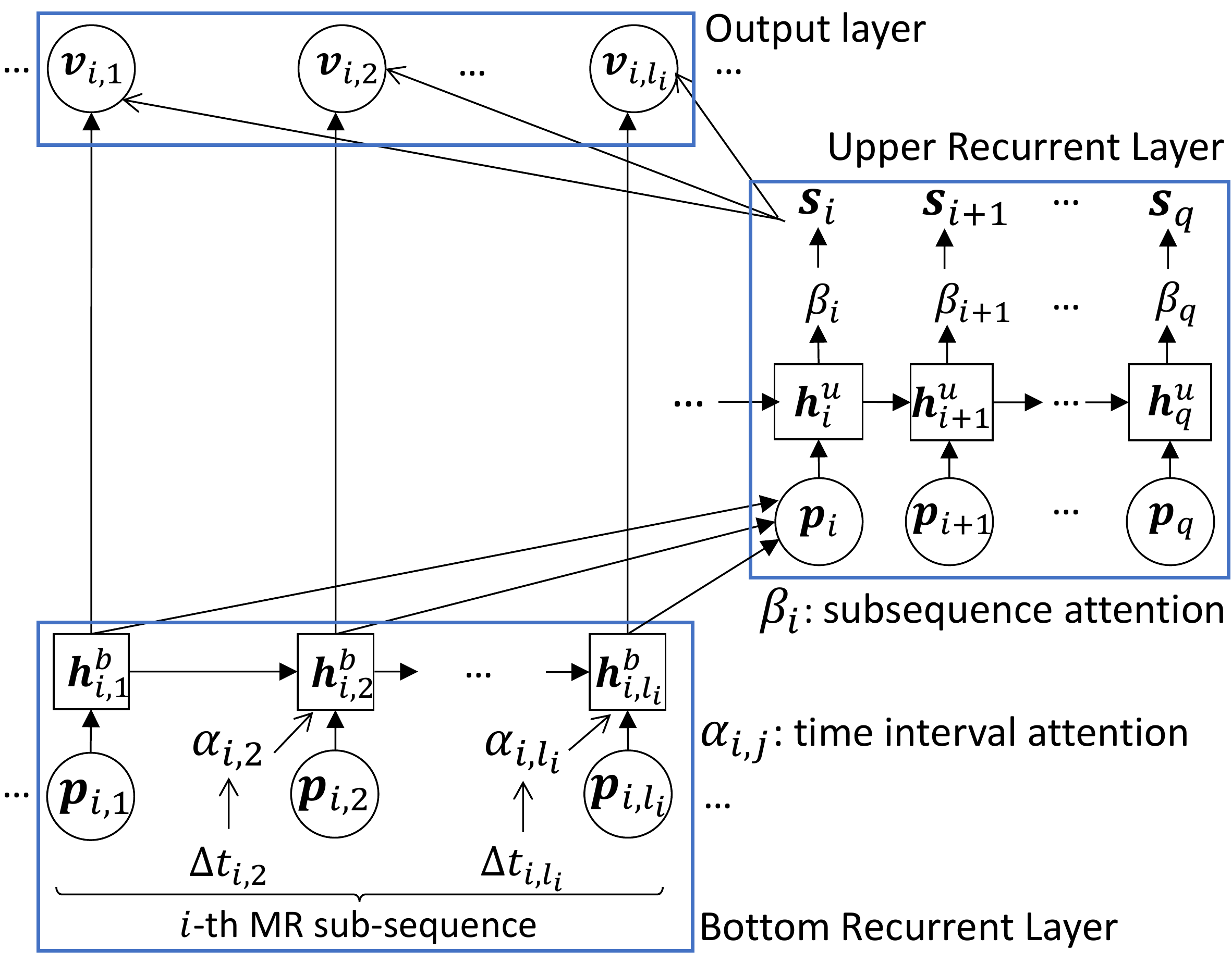}}\vspace{-2ex}
    \caption{Network Structure of Global Recurrent Predictor} \label{fig:global_componet}
\end{center}\vspace{-5ex}
\end{figure}

Figure \ref{fig:global_componet} gives the neural network structure of our global recurrent predictor. Inspired by the hierarchical deep neural network for document classification problem \cite{DBLP:conf/naacl/YangYDHSH16}, the global recurrent predictor first employs a \emph{bottom recurrent layer} with time interval attention to extract the short-term latent dependencies among MR samples within a subsequence. An \emph{upper recurrent layer} is then exploited to learn the long-term dependencies among MR subsequences to generate subsequence attention. Finally, an \emph{output layer} merges the hidden state of bottom recurrent layer and generated subsequence attention of upper recurrent layer to produce the final output of this module.

\subsubsection{Bottom Recurrent Layer}\label{sec4.2.1}

The bottom recurrent layer adopts LSTM cells to capture the short-term dependencies among the hidden vectors $\bm{p}_{i,1},...,\bm{p}_{i,|S_i^s|}$ within a certain subsequence $S_i^s$. However, a standard LSTM model ignores the difference of time intervals between the neighbouring cells. The length of time interval indicates the relevance from the previous cell to the current one. For instance, if the time interval increases, the contribution of previous cell becomes weak. Thus, we utilize the time-interval attention mechanism to address this issue.

\begin{figure}[th]
\vspace{-1ex}
\begin{center}					
    \centerline{\includegraphics[height=3.5cm]{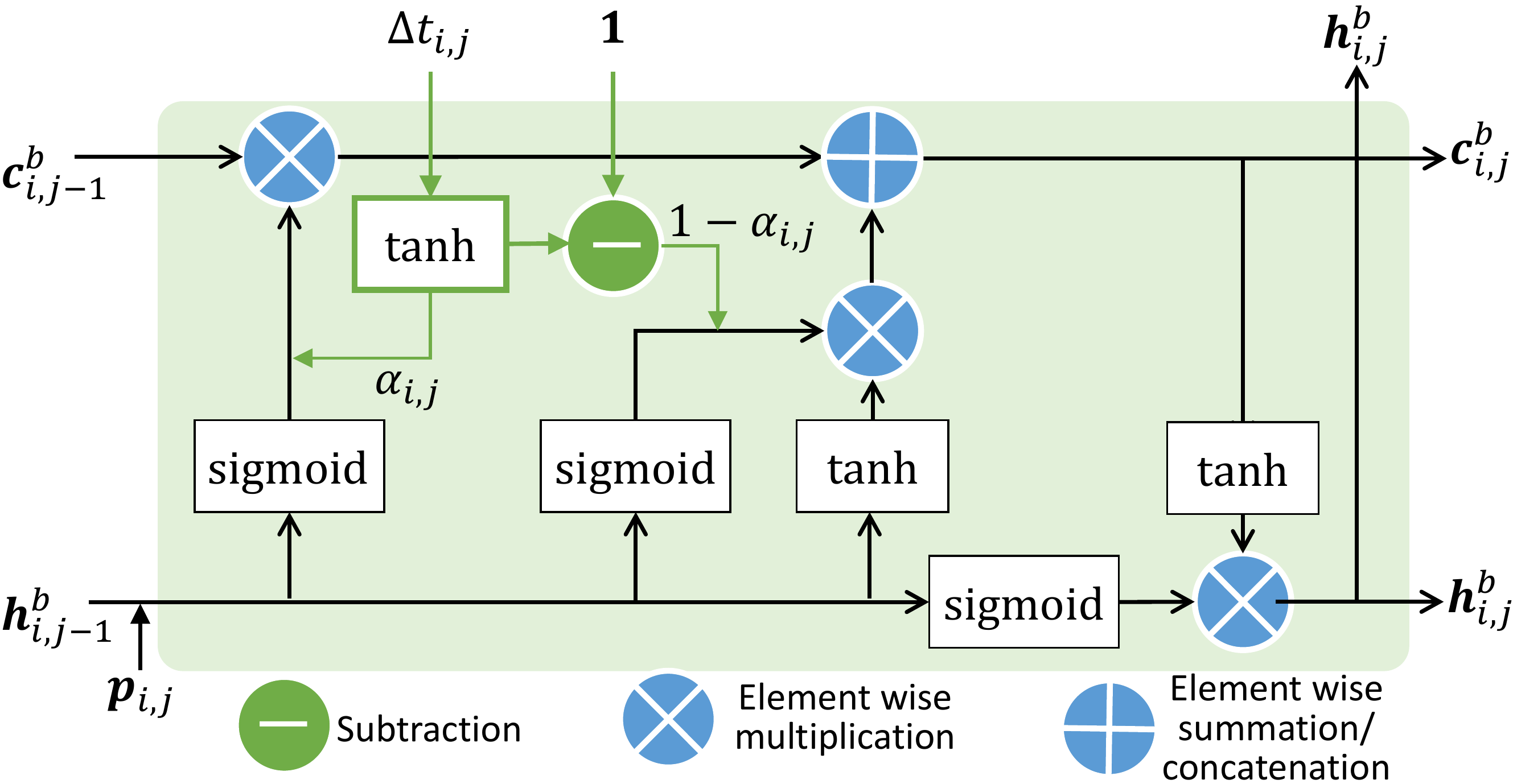}}\vspace{-2ex}
    \caption{Time-interval Attention in LSTM Cells} \label{fig:time_attn}
\end{center}\vspace{-2ex}
\end{figure}

\textbf{Time-interval attention}: Figure \ref{fig:time_attn} gives the LSTM cell structure with time interval attention. Specifically, the attention acts on the forget gate and input gate, respectively. We construct the time interval attention via a linear perceptron by referring to the time interval between the current state and the previous one with:
\begin{equation}\label{eq:aij}\small
\begin{aligned}
a_{i,j} &= \tanh(w^a \Delta t_{i,j}+b^a), \\
\end{aligned}
\end{equation}
where $w^a$ and $b^a$ are learnable parameters and $\Delta t_{i,j}$ is the timestamp difference between the current MR sample w.r.t $\bm{p}_{i,j}$ and the previous one. We denote the time interval attention weight by $a_{i,j}$ to update the forget and input gates of LSTM cell as:
\begin{equation}\small
\begin{aligned}
\bm{z}_{i,j,n}^{b} &= \tanh (\bm{W}_{z}^{b}[\bm{h}_{i,j,n-1}^{b};\bm{x}_{i,j,n}]+\bm{b}_{z}^{b}), \\
\bm{f}_{i,j,n}^{b} &= \sigma(\bm{W}_{f}^{b}[\bm{h}_{i,j,n-1}^{b};\bm{x}_{i,j,n}]+\bm{b}_{f}^{b})\cdot a_{i,j} \\
\bm{g}_{i,j,n}^{b} &= \sigma(\bm{W}_{g}^{b}[\bm{h}_{i,j,n-1}^{b};\bm{x}_{i,j,n}]+\bm{b}_{g}^{b})\cdot (1-a_{i,j}), \\
\bm{c}_{i,j,n}^{b} &= \bm{f}_{i,j,n}^{b} * \bm{c}_{i,j,n-1}^{b}+\bm{g}_{i,j,n}^{b} * \bm{z}_{i,j,n}^{b}, \\
\bm{o}_{i,j,n}^{b} &= \sigma (\bm{W}_{o}^{b}[\bm{h}_{i,j,n-1}^{b};\bm{x}_{i,j,n}]+\bm{b}_{o}^{b}), \\
\bm{h}_{i,j,n}^{b}&= \bm{o}_{i,j,n}^{b} * \tanh \bm{c}_{i,j,n}^{b},
\end{aligned}
\end{equation}
where superscript $b$ indicates the \underline{b}ottom recurrent layer. The time attention $a_{i,j}$ first acts on the forget gate and models a temporal decay to discard the information from the previous cell state. Next, since the latest state mainly determines the output of the current input, the attention $(1-a_{i,j})$ is thus applied to the input gate.


\subsubsection{Upper Recurrent Layer}
After the bottom recurrent layer has extracted the short-term temporal dependencies among the MR samples within a subsequence, the upper recurrent layer next captures the long-term dependencies between MR subsequences with the following input:
\begin{equation}\small
\begin{aligned}
\bm{p}_{i}= \sum_{j}{\bm{h}_{i,j}^{b}},
\end{aligned}
\end{equation}
where $\bm{h}_{i,j}^{b}$ denotes the $j$-th hidden state of the $i$-th subsequence acquired from the bottom recurrent layer. Since the upper recurrent layer is to capture the correlations among subsequences, the input of this layer, $\bm{p}_i$, needs to consider the latent characteristics of all elements in the $i$-th subsequence. Thus, we compute the input $\bm{p}_i$ by the sum of the hidden states in the $i$-th subsequence.

\textbf{Subsequence attention}: In the upper recurrent layer, we again leverage the LSTM cells to capture the long-term dependency by a subsequence attention mechanism to adaptively select meaningful hidden states (subsequences) across all subsequences ($i=1,...,q$). Specifically, the attention weight of the $i$-th hidden state $\bm{h}^u_i$ regarding the $i$-th subsequence (denoted by $\bm{p}_i$) is computed as:
\begin{equation}\small
\begin{aligned}
u_i &= (\bm{v}^u)^{\mathsf{T}}\tanh(\bm{W}^u \bm{h}_{i}^{u}+\bm{b}^u), \\
\beta_{i} &= \frac{\exp(u_i)}{\sum_{m=1}^{q}\exp(u_m)},
\end{aligned}
\end{equation}
where $\bm{v}^u$, $\bm{W}^u$ and $\bm{b}^u$ are learnable parameters. The attention weight $\beta_{i}$ represents the importance of the $i$-th hidden state $\bm{h}_{i}^{u}$ for the prediction, where superscript $u$ refers to the \underline{u}pper recurrent layer. Based on the weight $\beta_{i}$, the upper recurrent layer generates a weighted context vector $\bm{s}_{i}=\beta_{i}\bm{h}_{i}^{u}$ as the output.

\subsubsection{Output Layer}
The output layer is a fully connected (FC) layer to merge the outputs of the bottom and upper recurrent layer. The output layer takes the hidden state of bottom layer at time step $j$ of the $i$-th subsequence (denoted by $\bm{h}_{i,j}^{b}$) and the context vector of upper layer of the $i$-th subsequence (denoted by $\bm{s}_{i}$) as input. The final output of this module $\bm{v}_{i,j}\in\mathbb{R}^{d_{f}}$, where $d_f$ indicates the dimension of the learned feature vector, can be formulated as:
\begin{equation}\small
\begin{aligned}
\bm{v}_{i,j} = \tanh(\bm{W}^{o}_{b}\bm{h}_{i,j}^{u}+\bm{W}^{o}_{u}\bm{s}_{i}+\bm{b}^{o}),
\end{aligned}
\end{equation}
$\bm{W}^{o}_{b}$, $\bm{W}^{o}_{u}$ and $\bm{b}^o$ are learnable parameters of this layer, and the entire \textsf{Feature Extraction Module} can be treated as a shared encoder to learn common latent feature representation among different tasks automatically.

\section{Details of Two Learning Tasks}\label{s:task_modules}
In this section, we present the details of two learning tasks (outdoor position recovery and transportation mode detection) to process the learned features above.

\subsection{Outdoor Position Recovery Task}\label{sec:loc_task}
\begin{figure}[th]
\vspace{-1ex}
\begin{center}					
    \centerline{\includegraphics[height=4.2cm]{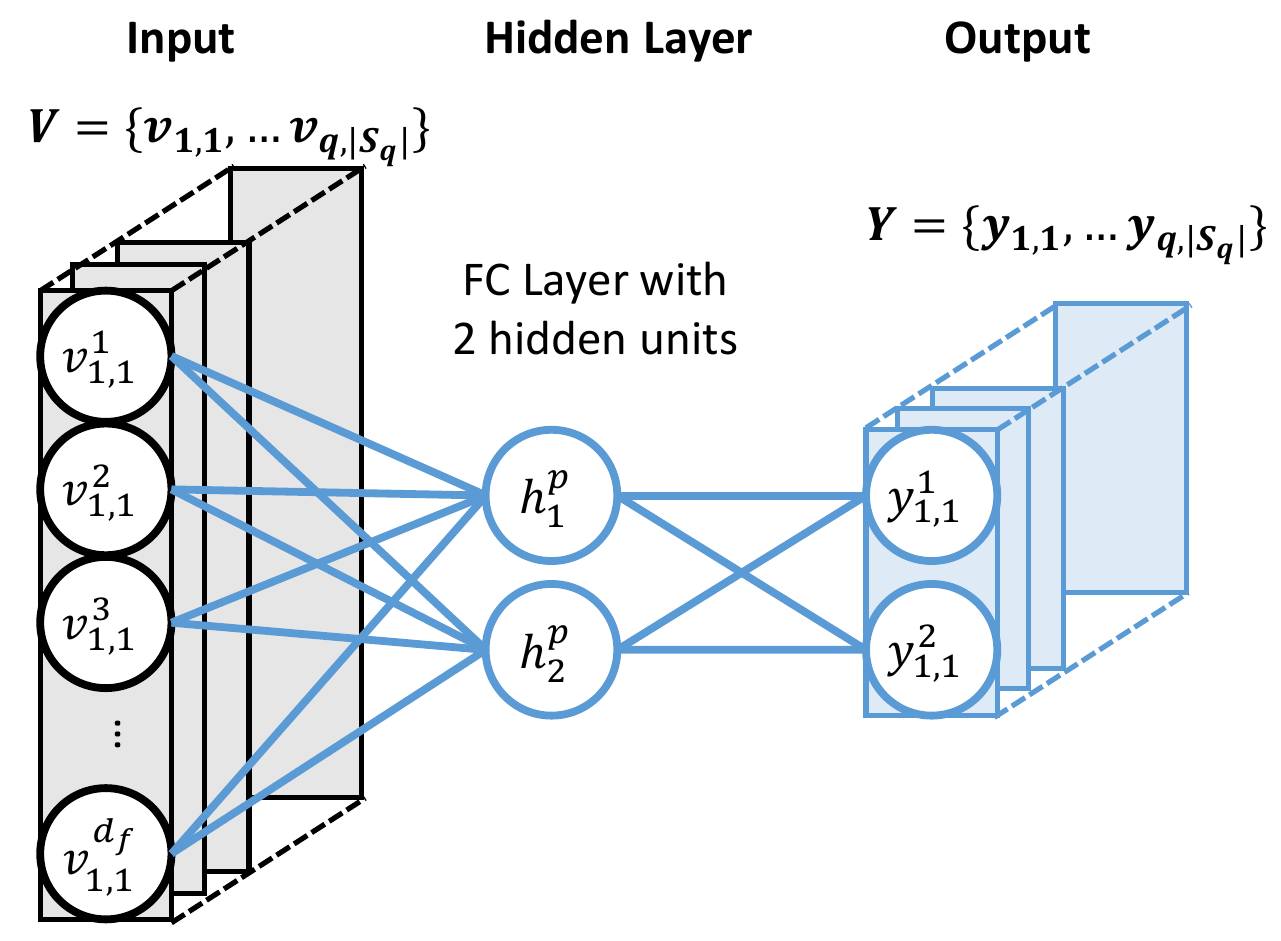}}\vspace{-2ex}
    \caption{Outdoor Position Recovery Task} \label{fig:loc_task}
\end{center}\vspace{-2ex}
\end{figure}

As shown in Figure \ref{fig:loc_task}, the outdoor position recovery task contains a Full Connection (FC) layer with two hidden units. Given the learned feature vector $\bm{V}$, the position recovery task generates a corresponding trajectory of MR positions $\bm{Y}$ where $\bm{y}_{i,j}\in \bm{Y}$ is a predicted GPS position of $v_{i,j}\in \bm{V}$. The output $\bm{y_{i,j}}$ can be formulated as:
\begin{equation}\small
\begin{aligned}
\bm{y}_{i,j} =\sigma(\bm{W}^{p}\bm{v}_{i,j}+\bm{b}^{p})
\end{aligned}
\end{equation}
where $\bm{W}^{p}$ and $\bm{b}^{p}$ are learnable parameters, and superscript $p$ indicates the task of position recovery. Due to the two dimensional GPS latitude and longitude coordinates, the output $\bm{y}_{i,j}$ is a $2 \times 1$ vector $\bm{y}_{i,j}$ with respect to one input MR sample $\bm{X}_{i,j}$ (or the learned vector $\bm{v}_{i,j}$ of $\bm{X}_{i,j}$). 

\subsection{Transportation Mode Detection Task}\label{sec:mode_task}

In {Figure \ref{fig:mode_task}, the transportation mode detection task still takes the shared feature vector $\bm{V}$ as input and generates a certain transportation mode for an input MR sample via a DNN classifier. The hidden layers of in the DNN classifier consist of $T$ stacked FC layers. Slightly different from  the position recovery task, the {transportation mode detection task} requires deeper network structure to perform the transportation mode estimation. In this task, the output $\bm{h}_{i,j,m}^{t}$ of the $m$-th hidden FC layer can be written as:
\begin{equation}\small
\left\{
\begin{aligned}
\bm{h}_{i,j,m}^{t} &=\sigma(\bm{W}^{t}_{m}\bm{v}_{i,j}+\bm{b}^{t}_{m}), \quad m=1 \\
\bm{h}_{i,j,m}^{t} &=\sigma(\bm{W}^{t}_{m}\bm{h}_{i,j,m-1}^{t}+\bm{b}^{t}_{m}), \quad m=\{2,...,T\} \\
\bm{m}_{i,j} &= Softmax(\bm{h}_{i,j,m}^{t}), \quad m=T
\end{aligned}
\right.
\end{equation}

\begin{figure}[th]
\vspace{-1ex}
\begin{center}					
    \centerline{\includegraphics[height=4.2cm]{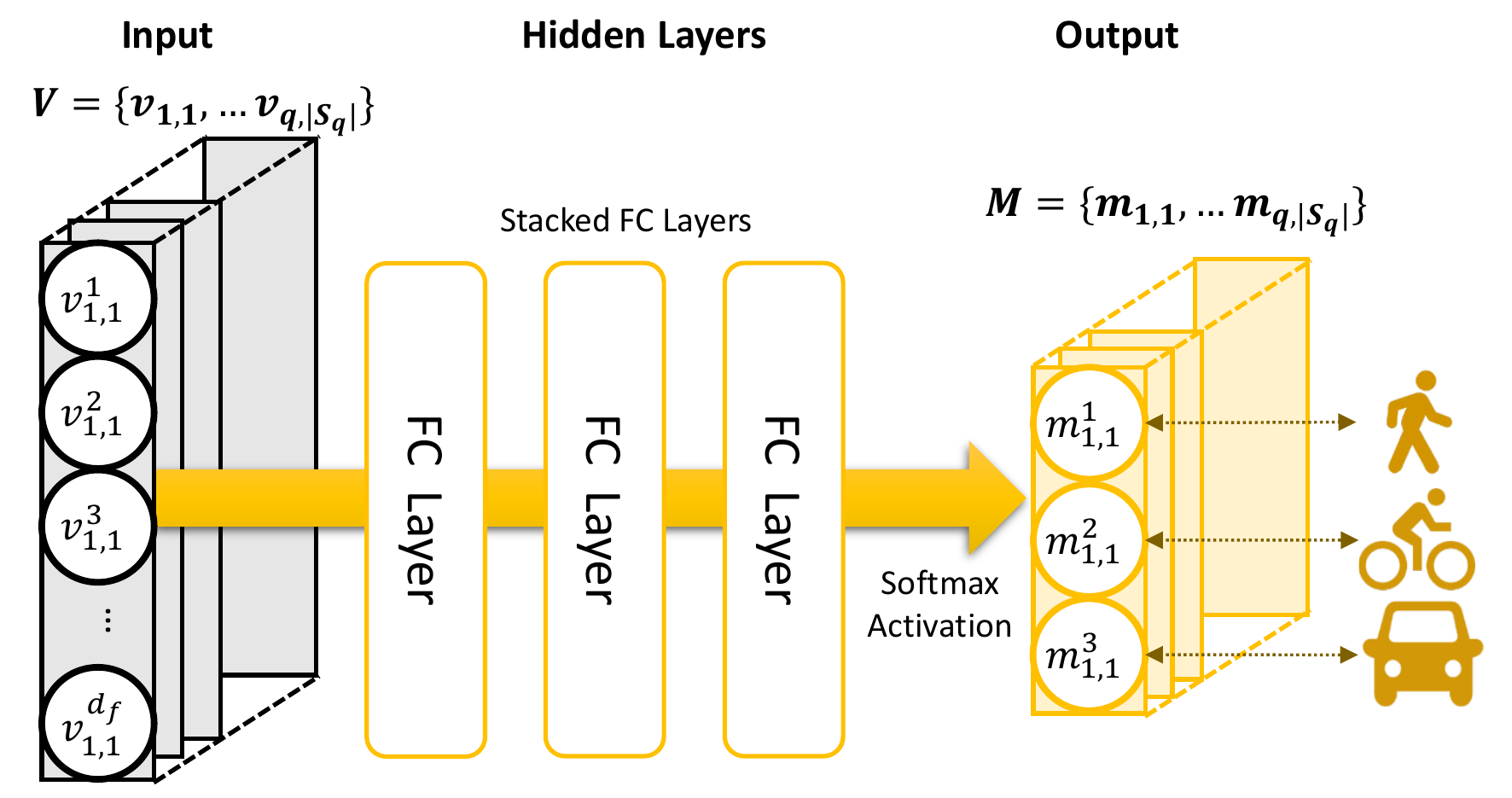}}\vspace{-2ex}
    \caption{Transportation Mode Detection Task} \label{fig:mode_task}
\end{center}\vspace{-2ex}
\end{figure}

where $\bm{W}^{t}_{m}$ and $\bm{b}^{t}_{m}$ are the learned parameters by FC layers, and superscript $t$ denotes the task of transportation mode detection. $\bm{m}_{i,j} \in \mathbb{R}^{|\mathcal{C}|}$ denotes the output of the final hidden layer with Softmax activation function, where $\mathcal{C}$ is a category set containing the possible transportation modes, e.g., walking, cycling, and driving in our datasets.} In this way, we classify a transportation mode $\bm{m}_{i,j}$ per MR sample, differing from the previous works Monitor \cite{DBLP:conf/vtc/Al-HusseinyY12}, and MonoSense \cite{DBLP:conf/vtc/AbdelAzizY15} which generate a certain mode for an entire window of MR samples. It is not hard to find that, in a fine-grained manner, our approach can process the heterogeneous MR samples with mixed transportation modes within an input MR (sub)sequence.


\section{Model Training}\label{s:prnet_training}
In this section, before giving the joint loss of the entire learning framework, we first introduce the individual loss involving of two learning tasks, and then present the training detail.
\subsection{Individual Loss}
First, since the outdoor position recovery problem is to predict numeric GPS coordinates, we model the problem as a regression task and compute the regression loss $\mathcal{L}^{\textbf{W}}_{loc}(S)$ with model parameters $\textbf{W}$ on input $S$ as:
\begin{equation}\small
\begin{aligned}
\mathcal{L}^{\textbf{W}}_{loc}(S) = \sum_{i=1}^{q}\sum_{j=1}^{|S_i|} ||\bm{y}_{i,j}-\bm{y}^{true}_{i,j}||
\end{aligned}
\end{equation}
where $\bm{y}^{true}_{i,j}$ is the ground truth location coordinates of the $j$-th input MR sample in the $i$-th subsequence $S_i$.

Second, we model the transportation mode detection problem as a classification task. Thus, to define the loss function $\mathcal{L}^{\textbf{W}}_{mode}$ of this task, we need to accumulate the mean cross-entropy for each MR sample in a certain MR sequence $S$ as follows
\begin{equation}\small
\begin{aligned}
\mathcal{L}^{\textbf{W}}_{mode}(S) &= \sum_{i=1}^{q}\sum_{j=1}^{|S_i|}\sum_{c=1}^{|\mathcal{C}|} \delta_{i, j}^{c}\log m_{i,j}^{c}
\end{aligned}
\end{equation}
In the equation above, $m_{i,j}^c$ denotes the predicted probability that the $j$-th input MR sample in the $i$-th subsequence $S_i$, and $\delta_{i, j}^{c}$ is the ground truth probability (0 or 1) of the input MR sample being the $c$-th transportation mode.

\begin{figure}[th]
\begin{center}					
    \centerline{\includegraphics[height=1.5in]{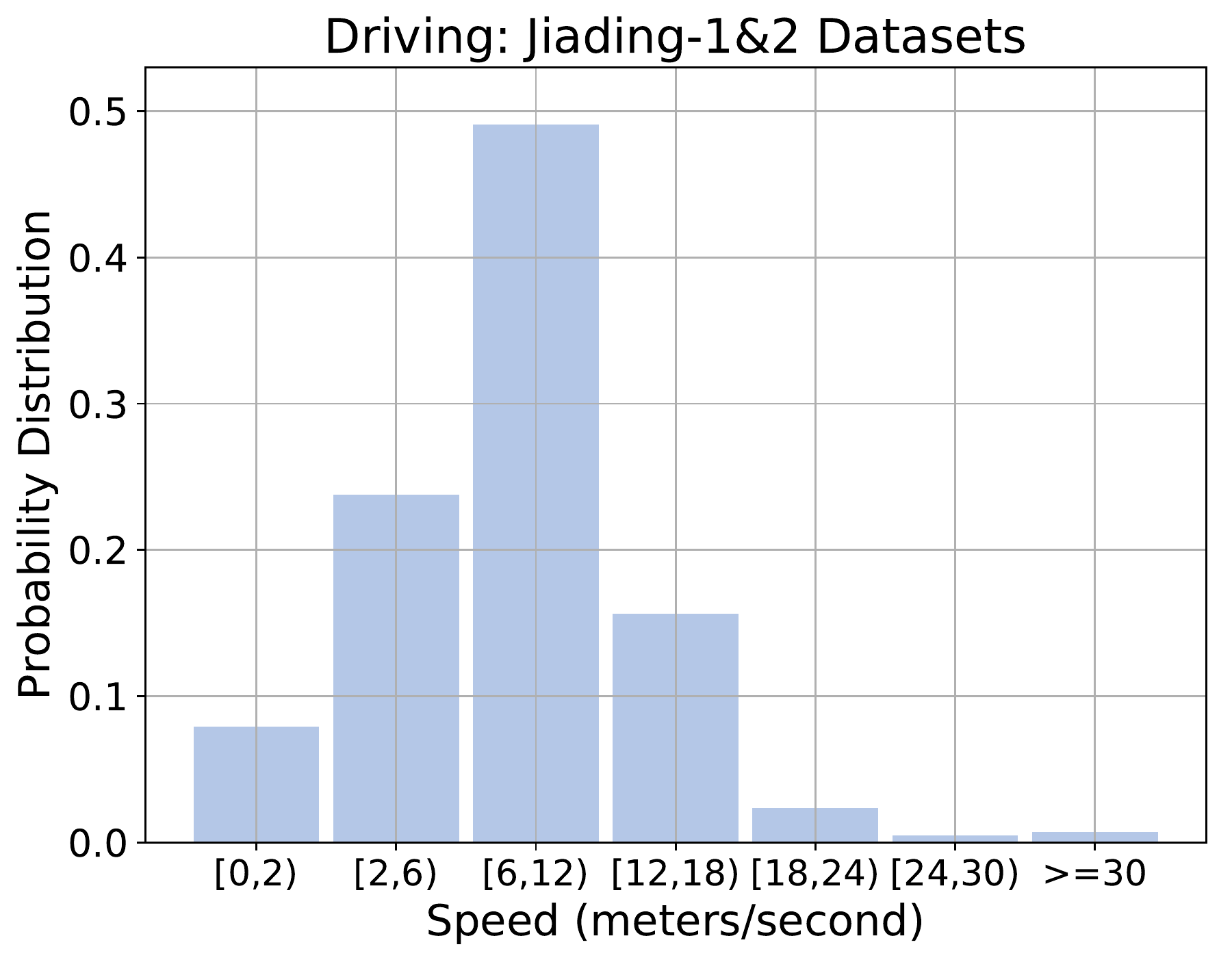}}
    \vspace{-2ex}
    \caption{Speed Probability Distribution of Driving Mode} \label{fig:speed_dis}
\end{center}\vspace{-2ex}
\end{figure}

Third, we note that the mobility features such as moving speed captured from predicted outdoor positions should be consist with the corresponding transportation mode. To this end, we introduce the speed constraint loss $\mathcal{L}^{\textbf{W}}_{speed}(\mathcal{S})$ to associate the recovered outdoor positions with detected transportation modes.
\begin{equation}\small
\begin{aligned}
\scriptsize
\mathcal{L}^{\textbf{W}}_{speed}(\mathcal{S}) &=\sum_{i=1}^{q}\sum_{j=2}^{|S_i|}- \log(1+P_{\bm{m}_{i,j}}(\hat{v}_{i,j})), \\ \hat{v}_{i,j}&=\frac{||\bm{y}_{i,j}-\bm{y}_{i,j-1}||}{t_{i,j}-t_{i,j-1}}
\end{aligned}
\end{equation}
In the speed-constraint loss above, the item $P_{\bm{m}_{i,j}}(\hat{v}_{i,j})$ indicates the likelihood that the speed $\hat{v}_{i,j}$ follows the transportation mode $\bm{m}_{i,j}$. Here, we estimate the speed $\hat{v}_{i,j}$ via the Euclidean distance $||\bm{y}_{i,j}-\bm{y}_{i,j-1}||$, between two predicted neighbouring positions $\bm{y}_{i,j}$ and $\bm{y}_{i,j-1}$. After that, given a speed probability distribution of a certain transportation mode, the item $P_{\bm{m}_{i,j}}(\hat{v}_{i,j})$ indicates the likelihood of the estimated speed $\hat{v}_{i,j}$ following such a transportation mode. For example, Figure \ref{fig:speed_dis} gives the speed probability distribution of the driving mode in one of our dataset. Given an estimated speed $\hat{v}_{i,j}=4m/s$, with the help of the  speed probability distribution, we then have the probability $P_{\bm{m}_{i,j}}(\hat{v}_{i,j})=0.238$, indicating that this speed $\hat{v}_{i,j}$ follows the driving mode by the probability $0.238$. From the equation above, if the predicted positions are more likely to follow the predicted transportation modes, we have a smaller negative loss $\mathcal{L}^{\textbf{W}}_{speed}$, which will next minimize our joint loss $\mathcal{L}^{\textbf{W}}$.

\subsection{Joint Loss}
A simple joint loss is to directly sum the three loss functions above. However, such a joint loss does not work well due to the significantly difference scale of the three loss functions. {To tackle this issue in a multi-task learning framework, we follow the previous work \cite{KendallGC18} to compute the weighted joint loss function via the so-called homoscedastic uncertainty to learn a weight for each task. Here, such an uncertainty is a quantity which stays constant for all input data and varies from tasks in Bayesian modelling.} 

Note that the two learning tasks in \textsf{PRNet$^+$} predict the outputs consisting of numeric GPS coordinates (i.e., continuous values) $\bm{Y}$ and a transportation mode (i.e., a discrete value) $\bm{M}$, respectively. Following the work \cite{KendallGC18}, we model such continuous and discrete values as a Gaussian likelihood and a Softmax likelihood, and thus have the following loss function with model parameters $\textbf{W}$ on input $S$ and two outputs $\bm{Y}$ and $\bm{M}$ for the two tasks:
\begin{equation}\small
\label{eq:joint-loss-1}
\begin{aligned}
\mathcal{L}^{\textbf{W}}(\mathcal{S}) =&-\log p(\bm{Y}, \bm{M}|\textbf{f}^{\textbf{W}})\\
 = & -\log\mathcal{N}(\bm{Y};\textbf{f}^{\textbf{W}},\sigma_1^2)\cdot Softmax(\bm{M};\textbf{f}^{\textbf{W}},\sigma_2^2)\\
 \approx & \frac{1}{2\sigma_1^{2}}\mathcal{L}^{\textbf{W}}_{loc}(\mathcal{S})+ \frac{1}{2\sigma_2^{2}}\mathcal{L}^{\textbf{W}}_{mode}(\mathcal{S}) +  \log\sigma_1 + \log\sigma_2
\end{aligned}
\end{equation}
where {$\textbf{f}^{\textbf{W}}$ denotes the sufficient statistic to derive a joint loss function by maximizing the Gaussian likelihood} and $\sigma_1$, $\sigma_2$ are learnable parameters which can be treated as the task weights \cite{KendallGC18}. 

Besides the two loss functions with respect to the two learning tasks, we have the speed-constraint loss and then incorporate the speed-constraint loss into the final joint loss as follows.
\begin{equation}\small
\label{eq:joint-loss}
\begin{aligned}
\mathcal{L}^{\textbf{W}}(\mathcal{S}) =& \frac{1}{2\sigma_1^{2}}\mathcal{L}^{\textbf{W}}_{loc}(\mathcal{S})+ \frac{1}{2\sigma_2^{2}}\mathcal{L}^{\textbf{W}}_{mode}(\mathcal{S}) + \\ & \alpha\mathcal{L}^{\textbf{W}}_{speed}(\mathcal{S}) + \log\sigma_1 + \log\sigma_2
\end{aligned}
\end{equation}
where $\alpha$ is a pre-defined hyper-parameter (e.g., $\alpha=0.05$ in one of our used datasets) for speed constraint in the joint loss. Note that we do not define a homoscedastic uncertainty weight for the speed-constraint loss $\mathcal{L}^{\textbf{W}}_{speed}(\mathcal{S})$. It is mainly because  $\mathcal{L}^{\textbf{W}}_{loc}(\mathcal{S})$ and $\mathcal{L}^{\textbf{W}}_{mode}(\mathcal{S})$ are those loss functions with respect to the two learning tasks with certain outputs and yet $\mathcal{L}^{\textbf{W}}_{speed}(\mathcal{S})$ not.

\subsection{Training Detail}
To train the two-task learning framework of \textsf{PRNet}${^{+}}$, we split the training data into mini-batches and pad MR sequences (both feature and label sequences) with a certain value e.g., -1. In this way, we make sure that they have the same length (mainly because the length of MR sequence with respect to IMSI differs in our datasets). In order to reduce the negative effect of the padding operation, we define a mask matrix $\mathcal{M}$ where the matrix member $\mathcal{M}_{i,j}$ is defined as:
\begin{equation}\small
\mathcal{M}_{i,j}=\left\{
\begin{aligned}
0 & , & \bm{y}_{i,j}^{true}=[-1, -1], \\
1 & , & \bm{y}_{i,j}^{true}\neq[-1, -1].
\end{aligned}
\right.
\end{equation}
where $\bm{y}_{i,j}^{true}=[-1, -1]$ indicates that $X_{i,j}$ is a padding item. Thus, we then update the aforementioned loss functions.
\begin{equation}\small
\begin{aligned}
\mathcal{L}^{\textbf{W}}_{mode}(S) &=\sum_{i=1}^{q}\sum_{j=1}^{|S_i|}\sum_{c=1}^{|\mathcal{C}|} \mathcal{M}_{i,j}\delta_{i, j}^{c}\log m_{i,j}^{c} \\
\mathcal{L}^{\textbf{W}}_{loc}(S) &=\sum_{i=1}^{q}\sum_{j=1}^{|S_i|}\mathcal{M}_{i,j} ||\bm{y}_{i,j}-\bm{y}^{true}_{i,j}||\\
\mathcal{L}^{\textbf{W}}_{cons}(\mathcal{S}) &=\sum_{i=1}^{q}\sum_{j=2}^{|S_i|}-\mathcal{M}_{i,j} \log(1+P_{\bm{m}_{i,j}}(\hat{v}_{i,j}))
\end{aligned}
\end{equation}

Until now, we give the training steps in Algorithm 1. Here, we take an Adam Optimizer to optimizing the model parameters.
\begin{algorithm}\small
  \caption{\textsf{PRNet}${^{+}}$ Training Procedure}
  \KwIn{Training set $\mathcal{D}$, Position recovery and mode detection ground truth: $\bm{Y}^{true}$ and $\bm{M}^{true}$, Training steps $T$, Batches $B$, Hyper-parameter $\alpha$}
  \KwOut{Model parameters}
  Initialize Model params $\textbf{W}$, $\sigma_1$, $\sigma_2$\;
  \For{$t = 1$ $\rightarrow$ $T$}
  {
   \For{$i = 1$ $\rightarrow$ $B$}
    {
      $\bm{V}_{b_i} \gets$ FeatureExtract($\mathcal{D}_{b_i}$)\;
      $\bm{Y}_{b_i} \gets$ PositionRecoveryTask($\bm{V}_{b_i}$)\;
      $\bm{M}_{b_i} \gets$ ModeDetectionTask($\bm{V}_{b_i}$)\;
      $\mathcal{L}^{\textbf{W}} \gets$ JointLoss($\bm{Y}_{b_i}, \bm{M}_{b_i}, \bm{Y}^{true}_{b_i}, \bm{M}^{true}_{b_i}, \alpha$)\;
      $\textbf{W}, \sigma_1, \sigma_2 \gets$ ADAM($\mathcal{L}^{\textbf{W}}, \textbf{W}, \sigma_1, \sigma_2$)\;
    }
 }
 Return $\textbf{W}$, $\sigma_1$ and  $\sigma_2$\;
\end{algorithm}\label{alg:train}

%% file: 04-eva.tex
\begin{table*}[!htp]\vspace{-2ex}
\scriptsize
\centering
\caption{Statistics of Used Data Sets}\label{tab:dataset}\vspace{-2ex}
\begin{tabular}{|l|c|c|c|c|c|}
\hline
	&\textbf{Jiading-1} & \textbf{Jiading-2} & \textbf{Jiading-3} &\textbf{Siping}&\textbf{Xuhui}\\
\hline
\hline
Num. of samples (2G/4G) &17354/12245 &8444/$-$ &$-$/150288& 6723/4953 & 14680/10455 \\
Route len (2G/4G) in km & 96.5/60.3& 62.35/$-$ &$-$/1347.6 & 24.6/15.5&29.3/15.9\\
Sampling rate (sec) & 3 & 1& 10 &3  &  2  \\
Area size (km$^2$) & 1.67 & 1.58 & 20.38 & 0.862 & 0.57 \\
\% of MRs with walking/cycling/driving &60.1/25.2/14.1 &58.8/31.1/10.1&unknown&63.8/36.2/- &58.5/-/41.5 \\
Num. of Serving BS per km$^2$ (2G/4G) & 26.34/29.43 &18.99/$-$ & $-$/24.92& 27.16/34.67 & 28.18/37.12 \\
Num. of serving BSs (2G/4G) & 62/38  &30/$-$ &$-$/508&51/72 & 28/16 \\\hline
\end{tabular}
\end{table*}

\section{Experiments}\label{sec:eva}
\subsection{Experimental Setup}
\textbf{1) Data Sets}: As shown in Table \ref{tab:dataset}, we evaluate \textsf{PRNet$^+$} using the data sets collected in three representative areas in Shanghai, China: a core business area \emph{Xuhui}, an urban area \emph{Siping} and a rural area \emph{Jiading}. The geographical distances between \emph{Jiading} and \emph{Siping}, between \emph{Jiading} and \emph{Xuhui}, and between \emph{Siping} and \emph{Xuhui} are around 31 km, 37 km and 15 km, respectively. We have three datasets in \emph{Jiading} and instead only one dataset in \emph{Siping} and \emph{Xuhui}, respectively. The \emph{Xuhui} and \emph{Jiading-3} datasets were provided by one of the largest Telco operators in China, and the data sets for \emph{Jiading-1$\sim$2} and \emph{Siping} were collected by our developed Android mobile app. When mobile users are moving around in outdoor road networks to collect MR samples, we switch on GPS receivers on Android mobile phones to collect current GPS coordinates. Since the collected GPS coordinates may contain noises, we further employ the map-matching technique  \cite{ZhangRDZXA18,HuangRZZYZ18} to mitigate the effect of noises. To protect user privacy, all sensitive information such as IMSI has been anonymized.

Table \ref{tab:dataset} summarizes the used data sets. For each dataset, we have collected 2G GSM and (or) 4G LTE MR samples and the associated GPS coordinates. Due to the limitations of the Android APIs in the 4G networks, each 4G MR sample in \emph{Jiading-1} and \emph{Siping} contains only one (serving) base station without the information about other neighboring base stations. Yet each 4G MR sample in \emph{Xuhui} and \emph{Jiading-3} still contains up to 7 base stations. The transportation mode ground truth of MR samples in \emph{Jiading-1$\sim$2}, \emph{Siping} and \emph{Xuhui} datasets has been labelled. However, the Telco operator does not provide the label information of transportation modes of \emph{Jiading-3} dataset due to the large size of \emph{Jiading-3} dataset.

The data sets contain heterogeneous MR samples due to mixed transportation modes and uncertain timestamp intervals between neighbouring samples. In particular, though we set a fixed sampling rate to collect MR data, the timestamp intervals between neighbouring samples vary significantly. For example, in the \emph{Jiading}-1 2G data set, the timestamp intervals vary from 1 to 125 seconds (see Table \ref{tab:timeinterval}), mainly due to (1) the uncertain delay of Android threads scheduled by Android OS to collect MR samples and (2) noisy MR samples (e.g., those samples having empty or zero signal measurements or empty base station IDs, and we have to clean the noise). Given the heterogeneous samples, we will study the effect of transportation modes and time intervals between neighboring MR samples in Section \ref{sec:sensitive}.

\begin{table}[!htp]\vspace{-2ex}
\scriptsize
\centering
\caption{Time Intervals in \emph{Jiading-1} 2G Dataset}\label{tab:timeinterval}\vspace{-2ex}
\begin{tabular}{|l|c|c|c|c|c|c|}
\hline
Intervals (s) &[1,3] & (3,5] & (5,10] & (10,30] & (30,60] & (60,125]\\
\hline
\hline
Ratio &71.32\% & 17.08\% & 8.14\% & 1.77\% & 1.34\% & 0.35\% \\\hline
\end{tabular}
\end{table}

Besides, we acquire the GPS longitude and latitude coordinates of every base station provided by Telco operators, and use them as additional features of MR samples. However, the parameters such as antenna height and angle are unavailable due to the limited information provided by the database. We believe that these parameters, if available, will further improve \textsf{PRNet$^+$}.

\begin{table}[hb]\vspace{-2ex}
\scriptsize
\centering
\caption{Counterparts}\label{tab:methods}\vspace{-2ex}
\begin{tabular}{|l|l|l|}
\hline
\textbf{Counterpart} & \textbf{Description} & \textbf{Approach} \\\hline\hline
\textsf{NBL} & Recent fingerprinting method \cite{MargoliesBBDJUV17} & Single Point \\ 
\textsf{DeepLoc} & 3-layer neural network \cite{DBLP:conf/gis/ShokryTY18} & Single Point \\
\textsf{RaF} & 1-layer Random Forest regression \cite{ZhuLYZZGDRZ16}& Single Point \\ %
\textsf{CCR} & 2-layer Random Forest regression \cite{ZhuLYZZGDRZ16}& Implicit Sequence \\\hline
\textsf{HMM}&  HMM + particle filtering \cite{RayDM16}& Sequence \\
\textsf{SeqtoSeq}& a LSTM-based seq. to seq. model \cite{DBLP:conf/nips/SutskeverVL14}& Sequence \\
\textsf{ConvLSTM} & a convolutional LSTM \cite{DBLP:conf/nips/ShiCWYWW15} & Sequence \\
\textsf{PRNet} & a hierarchical neural network \cite{DBLP:conf/cikm/ZhangRZYZ19} & Sequence \\
\textsf{PRNet$^+$} & an improvement on \textsf{PRNet} & Sequence \\\hline
\end{tabular}
\end{table}

\textbf{2) Counterparts}: We compare \textsf{PRNet}$^+$ against 8 counterparts in Table \ref{tab:methods} from the following aspects:

\emph{(a)} Depending upon the location recovery result, the counterparts are either single point- (\textsf{NBL} \cite{MargoliesBBDJUV17}, \textsf{RaF} \cite{ZhuLYZZGDRZ16}, \textsf{CCR} \cite{ZhuLYZZGDRZ16} and a very recent work \textsf{DeepLoc} \cite{DBLP:conf/gis/ShokryTY18}), or sequence-based approaches (\textsf{HMM} \cite{RayDM16}, \textsf{SeqtoSeq} \cite{DBLP:conf/nips/SutskeverVL14}, \textsf{ConvLSTM} \cite{DBLP:conf/nips/ShiCWYWW15}, \textsf{PRNet} \cite{DBLP:conf/cikm/ZhangRZYZ19} and \textsf{PRNet}$^+$). In \cite{ZhuLYZZGDRZ16}, the location recovery model can be either only a single-layer Random Forest (\textsf{RaF}), or a two-layer Random Forests (\textsf{CCR}) which can be treated as an implicit sequence-based approach due to the contextual features acquired from the predicted result of the 1st layer.

\emph{(b)} In terms of the used models, the counterparts can be either the fingerprinting-based (\textsf{NBL}), or traditional machine learning-based (\textsf{RaF}, \textsf{CCR} and \textsf{HMM}), or DNN-based models (\textsf{SeqtoSeq}, \textsf{ConvLSTM}, \textsf{DeepLoc}, \textsf{PRNet} and \textsf{PRNet}$^+$).

Given the MR samples above, we follow the previous work \cite{DBLP:journals/titb/HanninkKPBSGKE18, DBLP:journals/lgrs/ScottMDN17} to adopt the $k$-fold ($k=10$) cross validation by choosing 80\% training and 20\% testing data from each data set to avoid over-fitting. We compute the prediction error by the Euclidian distance between recovered locations and ground truth (i.e., the real GPS coordinates of MR samples), we choose median error, mean error, and top $90\%$ error (by sorting prediction errors in ascending order) as evaluation metrics. Since the transportation modes of \emph{Jiading-3} dataset are unavailable, we exploit the previous transportation mode detection approach on GPS data \cite{DBLP:journals/tweb/ZhengCLXM10} to predict the transportation modes of training MR samples in \emph{Jiading-3} dataset, and then the training MR samples of \emph{Jiading-3} dataset with the predicted transportation modes are together used to train \textsf{PRNet}$^+$ module for position recovery.

\textbf{3) Key Parameters}: Table \ref{tab:params} lists the parameters used in our experiments. We use default values in the baseline experiment, and vary the values within allowable ranges for multi-task learning and sensitivity study. Due to the low sampling rate and short length of {user trajectories} in \emph{Jiading-3} dataset, we thus treat an entire user trajectory as an MR sequence.

\begin{table}[!hbp]
\scriptsize
\centering
\caption{Key Parameters}\label{tab:params}\vspace{-2ex}
\begin{tabular}{|l|l|l|}
\hline
\textbf{Parameter}               & \textbf{Range}               & \textbf{Default Val.}             \\\hline\hline
Parameter $\alpha$ in Joint Loss   & 0.01, 0.05, 0.1   & 0.05     \\
Timestamp Interval (s)  & 3--120   & 3                     \\ 
Base Station Density    & 25\% - 100\%             & 100\%                 \\ 
Learning Rate of \textsf{PRNet$^+$}         &0.005, 0.001, 0.0005, 0.0001 & 0.0005\\ \hline
\end{tabular}
\end{table}

\subsection{Baseline Study}\label{sec:baseline}
\begin{table*}[tp]\vspace{-2ex}
\centering\scriptsize
\caption{Baseline Experiment: Localization Errors of Nine Approaches}\vspace{-2ex}
\label{tab:performance_comparison}
\begin{tabular}{|cc|cccc|ccccc|}
\hline
\multicolumn{2}{|c|}{\textbf{Methods}} & \textsf{NBL} & \textsf{DeepLoc} & \textsf{RaF} & \textsf{CCR} & \textsf{HMM} & \textsf{SeqtoSeq} & \textsf{ConvLSTM} & \textsf{PRNet} & \textsf{PRNet}$^+$ \\ \hline\hline
\multirow{3}{*}{\textbf{Jiading-1 (2G)}} & Median & 53.4 & 35.1 & 38.3 & 34.1 & 36.5 & 25.4 & 28.5 & 15.8 & 15.2 \\
 & Mean & 67.2 & 47.6 & 48.3 & 43.2 & 52.3 & 50.6 & 59.3 & 37.8 &  34.3\\
 & 90\% & 300.9 & 250.3 & 168.9 & 142.3 & 172.8 & 85.3 & 129.3 & 63.2 &  59.6\\ 
\multirow{3}{*}{\textbf{Jiading-1 (4G)}} & Median & 59.7 & 40.2 & 38.5 & 30.2 & 42.1 & 24.2 & 27.3 & 18.4 & 18.1 \\
 & Mean & 72.3 & 53.9 & 47.2 & 44.5 & 53.6 & 50.1 & 58.1 & 40.6 & 37.4 \\
 & 90\% & 318.6 & 280.6 & 158.9 & 145.9 & 188.4 & 81.7 & 124.5 & 66.5 & 62.3 \\ \hline
\multirow{3}{*}{\textbf{Jiading-2 (2G)}} & Median & 54.3 & 36.2 & 37.5 & 36.1 & 37.3 & 27.3 & 29.6 & 19.4 & 17.3 \\
 & Mean & 68.2 & 49.3 & 48.9 & 44.0 & 52.4 & 49.5 & 57.6 & 41.3 & 37.9\\
 & 90\% & 291.4 & 255.7 & 166.4 & 147.8 & 168.2 & 88.4 & 127.6 & 70.4 & 64.7 \\ \hline
\multirow{3}{*}{\textbf{Jiading-3 (4G)}} & Median & 55.3 & 44.7 & 44.3 & 40.3 & 51.8 & 33.7 & 34.5 & 31.5 & 28.9 \\
 & Mean & 64.7 & 54.2 & 53.6 & 51.1 & 56.2 & 54.1 & 57.4 & 49.5 & 47.8 \\
 & 90\% & 296.6 & 247.0 & 166.9 & 154.2  & 167.0 & 157.1 & 149.1 & 135.2 & 131.7  \\ \hline
\multirow{3}{*}{\textbf{Siping (2G)}} & Median & 42.8 & 33.5 & 41.1 & 37.5 & 37.2 & 23.4 & 27.7 & 15.3 & 14.9 \\
 & Mean & 63.0 & 44.7 & 47.3 & 42.8 & 51.4 & 48.9 & 57.3 & 34.2 & 32.3 \\
 & 90\% & 298.3 & 219.9 & 158.8 & 139.5 & 160.3 & 84.9 & 120.4 & 60.2 & 58.4 \\ 
\multirow{3}{*}{\textbf{Siping (4G)}} & Median & 43.2 & 38.7 & 35.1 & 29.5 & 30.7 & 22.7 & 25.6 & 17.4 & 17.2\\
 & Mean & 64.9 & 49.6 & 44.4 & 40.6 & 48.2 & 47.3 & 55.7 & 37.7 & 35.3 \\
 & 90\% & 256.7 & 267.5 & 105.8 & 101.2 & 145.6 & 83.5 & 117.2 & 63.4 & 60.1 \\ \hline
\multirow{3}{*}{\textbf{Xuhui (2G)}} & Median & 45.9 & 31.2 & 35.6 & 30.0 & 35.3 & 22.5 & 25.3 & 15.7 & 15.2 \\
 & Mean & 59.0 & 40.5 & 43.4 & 40.2 & 42.7 & 43.2 & 52.5 & 34.4 & 31.1 \\
 & 90\% & 240.7 & 210.5 & 137.7 & 125.2 & 148.0 & 84.7 & 113.2 & 62.5 & 58.4 \\ 
\multirow{3}{*}{\textbf{Xuhui (4G)}} & Median & 32.2 & 27.4 & 32.2 & 20.0 & 28.6 & 21.9 & 24.4 & 13.7 & 13.4 \\
 & Mean & 52.7 & 400.1 & 41.5 & 34.1 & 39.1 & 42.9 & 50.6 & 31.2 & 28.3 \\
 & 90\% & 191.2 & 211.9 & 104.4 & 98.3 & 129.5 & 81.8 & 107.8 & 59.3 & 57.1 \\ \hline
\end{tabular}
\end{table*}
In Table \ref{tab:performance_comparison}, we compare the prediction errors (median, mean, and 90\% errors) of the nine approaches on the eight datasets from three areas (Jiading, Siping, and Xuhui). From Table \ref{tab:performance_comparison},  we have the following findings.


\emph{1}) \textsf{PRNet$^+$} offers the least error among these nine approaches on all MR datasets. For example, in \emph{Jiading-1} 2G data set, \textsf{PRNet$^+$} reduces the median error by 56.7\% compared with the recent outdoor Telco position recovery approach, \textsf{DeepLoc}. It is mainly because \textsf{DeepLoc} is a single point-based prediction method, which does not capture contextual dependencies. In addition, \textsf{PRNet}$^+$ outperforms the two deep neural network-based approaches (i.e. \textsf{SeqtoSeq} and \textsf{ConvLSTM}). Such results indicate that \textsf{PRNet$^+$} outperforms both the state-of-the-arts and alternative DNN-based approaches.

\emph{2}) We compare the top-90\% errors among the nine approaches. Still on \emph{Jiading-1} 2G dataset, \textsf{PRNet$^+$} outperforms \textsf{CCR} with 58.1\% lower top-90\% error. Similar results are applicable to other datasets. These numbers indicate that \textsf{PRNet$^+$} has fully employed the power of hierarchical deep neural network, sequence model, attention mechanism and joint loss together to effectively mitigate the outliers among predicted locations.


\emph{3}) We compare the results of the five sequence-based approaches against four single-point-based ones. For example, a simple sequence model, \textsf{HMM}, still outperforms \textsf{DeepLoc} by 31.0\% smaller 90\% error on \emph{Jiading-1} 2G data set. In addition, though \textsf{CCR} and \textsf{RaF} are both Random Forest-based approaches, \textsf{CCR} leverages the contextual features such as moving speed and leads to better result than \textsf{RaF} by 10.9\% smaller median error. Such results verify that the sequence models, no matter they implicit or explicit use of contextual features, could lead to better results than the original single-point-based approaches.

\emph{4}) Among the five deep neural network-based approaches, \textsf{ConvLSTM} suffers the highest errors. It might be mainly because \textsf{ConvLSTM} is typically used to solve flow prediction (e.g., precipitation nowcasting  \cite{DBLP:conf/nips/ShiCWYWW15} and traffic accident \cite{DBLP:conf/kdd/YuanZY18}) for the time series data involving explicit spatio-temporal information. However, MR samples do not contain accurate locations, which are just our objective (note that the GPS locations of base stations are used as additional features, but not the locations of MR samples). Thus, \textsf{ConvLSTM} may not properly solve our problem though \textsf{ConvLSTM} also utilizes the power of both \textsf{CNN} and \textsf{LSTM}. \textsf{SeqtoSeq} is slightly better than \textsf{ConvLSTM}, but still incomparable with \textsf{PRNet$^+$}.

\emph{5}) Table \ref{tab:performance_comparison} is consistent with results reported in the previous works \textsf{CCR} and \textsf{NBL}: for the same area e.g., \emph{Xuhui}, 4G MR samples typically lead to better result than 2G samples; and for the same Telco networks, the results of the \emph{Xuhui} 2G dataset are better than those of \emph{Siping} and \emph{Jiading-1$\sim$2} 2G datasets. We have the finding that higher base station deployment density in 4G networks and core areas (see Table \ref{tab:dataset}) can achieve better result than those in 2G and rural areas. We also note that due to only one base station per MR sample in \emph{Jiading-1} and \emph{Siping} 4G datasets, their errors are even higher than those 2G datasets. Such result verifies the importance of multiple base stations for precise location recovery.

\emph{6)} Among the datasets in \emph{Jiading} area, the localization errors of \emph{Jiading-3} 4G data set are higher than the other datasets. It is mainly caused by the lowest sampling rate among all datasets and more sparse base station density within the area covered by \emph{Jiading-3} data set. In addition, the localization errors of \emph{Jiading-2} 2G data set are slightly higher than \emph{Jiading-1} 2G data set. It is mainly because \emph{Jiading-2} data set has a higher percentage of MR samples collected by driving and cycling (see Table \ref{tab:dataset}). We will discuss the effect of transportation modes in Section \ref{sec:sensitive}.

Due to space limitation, we choose \emph{Jiading-1} and \emph{Jiadng-2} 2G datasets (the two datasets contain the greatest number of MR samples labelled by transportation modes) to evaluate \textsf{PRNet$^+$} in the rest of this section.

\subsection{Evaluation of Multi-Task Learning Framework}\label{sec:mtl_eva}

In this section, we are interested in how the two individual learning tasks independently work against the entire multi-task learning framework of \textsf{PRNet$^+$}. 
In Table \ref{tab:mtl}, the first two rows indicate the performance result of an individual task alone, the third row involves those of the multi-task learning models with man-made uniform weights in the joint loss, and the rest rows give those with task uncertainty weighted loss functions.

\begin{table*}[tp]
\centering\scriptsize
\caption{Multi-Task Learning in PRNet${^+}$.}
\label{tab:mtl}
\begin{tabular}{|c|c|c|c|c|c|c|c|}
\hline
\multirow{3}{*}{Loss} & \multicolumn{3}{c|}{\multirow{2}{*}{Task (Loss item) Weights}} & \multicolumn{2}{c|}{Jiading-1 2G Data} & \multicolumn{2}{c|}{Jiading-2 2G Data} \\ \cline{5-8}
 & \multicolumn{3}{c|}{} & Pos. Recover. & Mode Detect. & Pos. Recover. & Mode Detect. \\ \cline{2-8}
 & Speed Const. & Pos. Recover. & Mode Detect. & 50\% Error: meters & Accuracy & 50\% Error: meters & Accuracy \\ \hline
\hline
Pos. Recovery Task alone & / & 1 & 0 & 15.8 & / & 16.7 & / \\
Mode Detect. Task alone & / & 0 & 1 & / & 0.754 & / & 0.761 \\ 
Uniform weights & 1 & 1 & 1 & 28.7 & 0.682 & 31.4 & 0. 679\\
2-task uncertainty weights & / & $\surd$ & $\surd$ & 17.1 & 0.775 & 17.9 & 0.752 \\
2-task uncertainty weights & 0.01 & $\surd$ & $\surd$ & 16.8 & 0.781 & 17.2 & 0.760 \\
2-task uncertainty weights & 0.1 & $\surd$ & $\surd$ & 17.4 & 0.781 & 18.1 & 0.766 \\ 
2-task uncertainty weights & 0.05 & $\surd$ & $\surd$ & \textbf{15.4} & \textbf{0.784} & \textbf{16.2} & \textbf{0.773} \\
\hline
\end{tabular}
\end{table*}

Table \ref{tab:mtl} clearly illustrates the benefit of the multi-task learning with much better performing results than an individual task alone. For example, using the multi-task learning framework, \textsf{PRNet$^+$} in the bottom row improves the classification accuracy from 75.4\% to 78.4\% in \emph{Jiading-1} 2G dataset and the position recovery error (median error) from 15.8 meters to 15.2 meters. In addition, we compare the uncertainty weights against the manual weights in the joint loss. Using a uniform weighting results in rather poor performance, even worse than the results of an individual task alone. Next, the 4-th row (i.e., the loss regarding to the speed constraint is missed in the joint loss) cannot compete the bottom row, i.e., the 2-task uncertainty weights with the pre-defined parameter $\alpha=0.05$ for the speed constraint loss. It is mainly because using the speed constraint can effectively link the predicted positions with detected modes for better results.



\begin{figure*}[th]\vspace{-2ex}
\hspace{-8ex}
\begin{center}
\begin{tabular}{c c c}

\begin{minipage}[t]{0.31\linewidth}
\begin{center}
\centerline{\subfigure{
\label{fig:trans:data}
\includegraphics[height=1.4in]{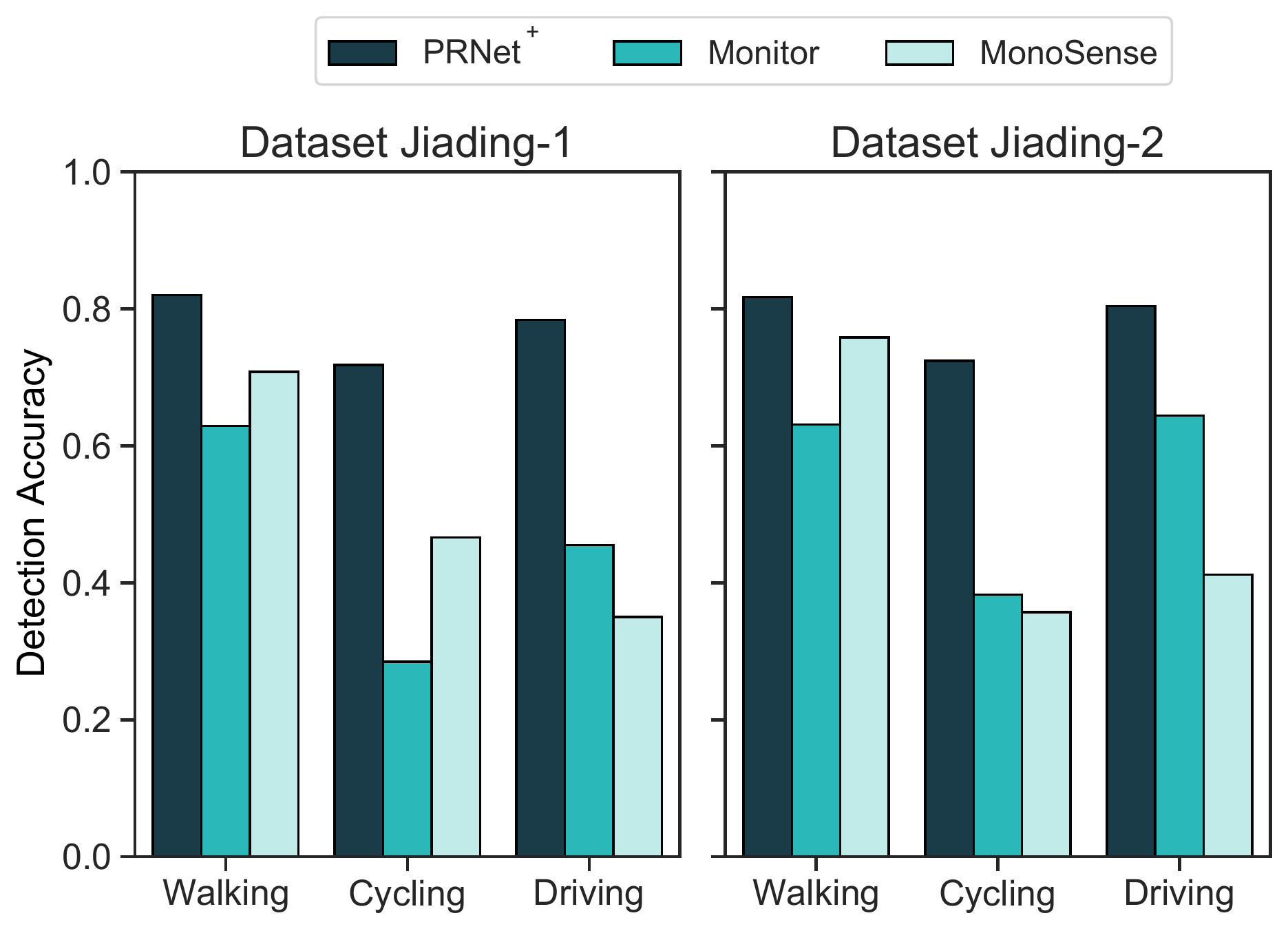}}}
\end{center}
\end{minipage}
\begin{minipage}[t]{0.31\linewidth}
\begin{center}
\centerline{\subfigure{
\label{fig:trans:improv}
\includegraphics[height=1.4in]{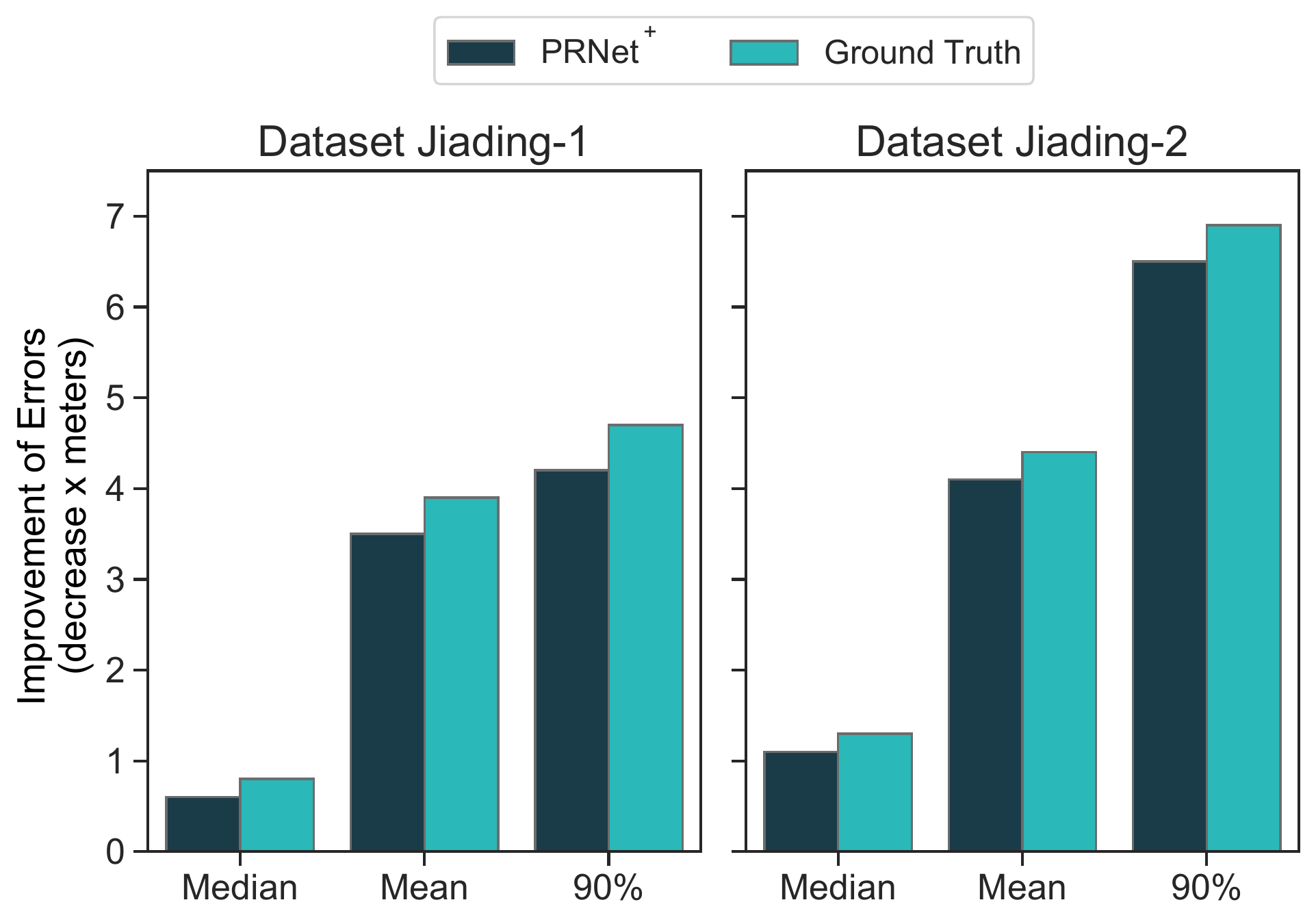}}}
\end{center}
\end{minipage}
&
\begin{minipage}[t]{0.31\linewidth}
\begin{center}
\centerline{\subfigure{
\label{fig:trans:seq}
\includegraphics[height=1.4in]{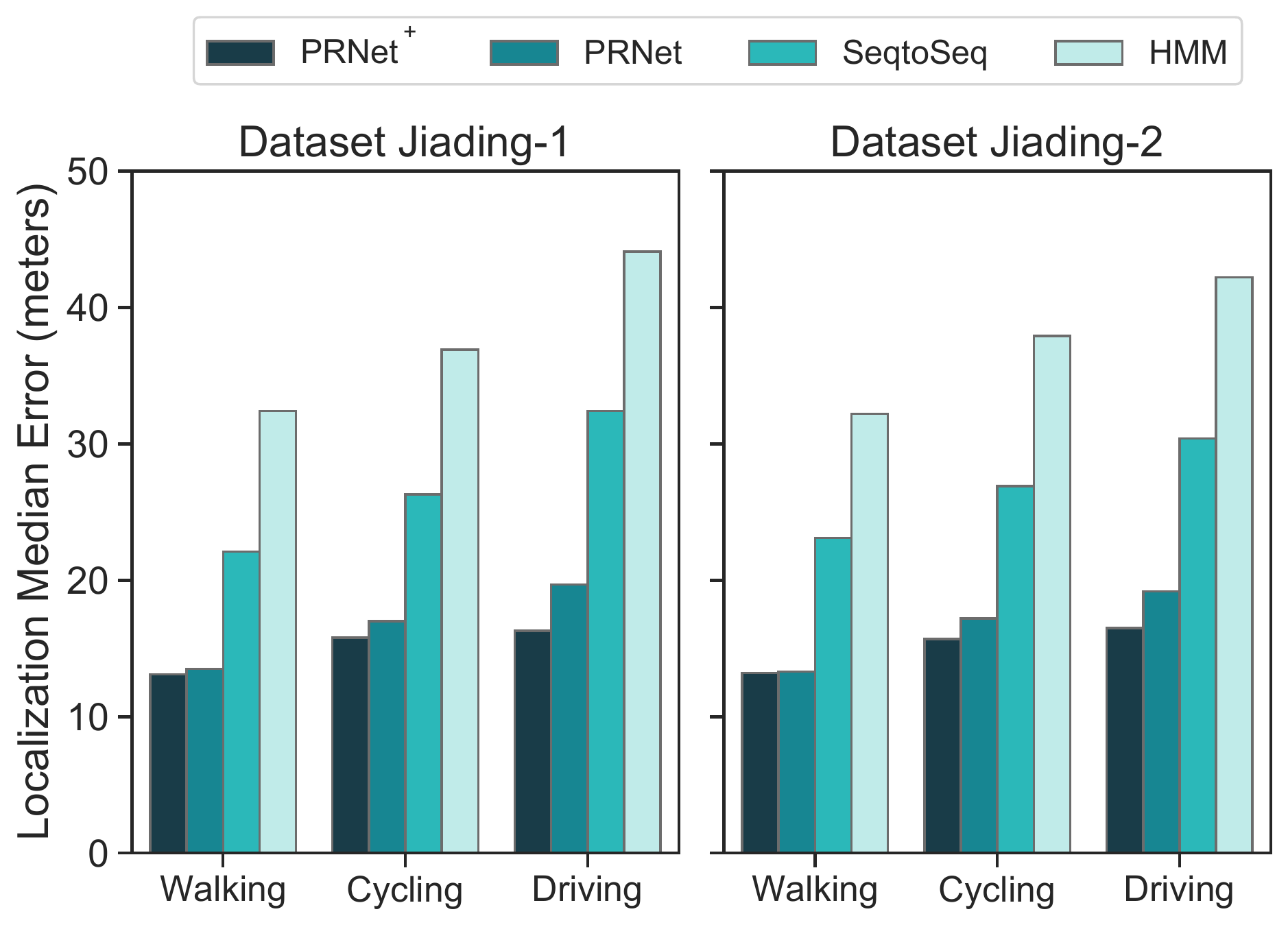}}}
\end{center}
\end{minipage}

\end{tabular}\vspace{-4ex}
\caption{Transportation Mode Detection: (a) Counterparts (b) Localization Improvement, (c) Sequence Model (from left to right).} \label{exp:trans_mode_method}\vspace{-2ex}
\end{center}
\end{figure*}

\subsection{Sensitivity Study}\label{sec:sensitive}
\noindent \textbf{(1) \underline{Transportation Mode}}: Due to the mixed transportation modes of MR samples, we are interested in how these modes affect the performance of \textsf{PRNet$^{+}$}. 

Firstly, Figure \ref{fig:trans:data} evaluates the detection accuracy of \textsf{PRNet$^+$} with two other approaches \textsf{Monitor} \cite{DBLP:conf/vtc/Al-HusseinyY12} and \textsf{MonoSense} \cite{DBLP:conf/vtc/AbdelAzizY15}. From this figure, we find that \textsf{PRNet$^+$} performs consistently better than \textsf{Monitor} and \textsf{MonoSense}. Since \textsf{Monitor} and \textsf{MonoSense} predict a certain transportation mode for entire MR samples within a time window, they do not work well when such MR samples involves mixed transportation modes within a time window. In addition, the detection accuracy of either walking or driving mode is significantly higher than the one of cycling. It is mainly because in our datasets collected in a campus, fast (resp. slow) cycling speed is close to the driving (resp. walking) speed, thus blurring the features w.r.t the cycling mode from those of either driving or walking modes.


Secondly, Figure \ref{fig:trans:improv} studies the benefit of transportation mode detection on position recovery. Specifically, we calculate how much localization error reduction is gained when the speed constraint loss is incorporated into the joint loss of \textsf{PRNet$^+$}. We find that the reduction of the mean and 90\% errors of \textsf{PRNet}$^+$ is rather significant. Such result indicates that transportation mode detection greatly improves the outliers of predicted positions.

Finally, under various transportation modes, Figure \ref{fig:trans:seq} compare \textsf{PRNet$^+$} against three other sequenced-based approaches \textsf{PRNet}, \textsf{SeqtoSeq} and \textsf{HMM}). From this figure, we find that the walking mode leads to the best prediction accuracy. It is mainly because the MR samples in the walking mode exhibit higher spatial locality compared to cycling and driving, i.e., more MR samples within every MR (sub)sequence. Moreover, among all transportation modes, \textsf{PRNet$^+$} still performs best. In particular, the errors of both \textsf{SeqtoSeq} and \textsf{HMM} grow significantly on the MR samples collected by the driving mode. When these MR samples are rather spare in terms of their locations and uneven timestamp intervals, these two models are hard to precisely capture spatio-temporal locality.



\begin{figure*}
\begin{center}
\begin{tabular}{c c c}
\begin{minipage}[t]{0.31\linewidth}
\begin{center}
\centerline{\includegraphics[height=1.4in]{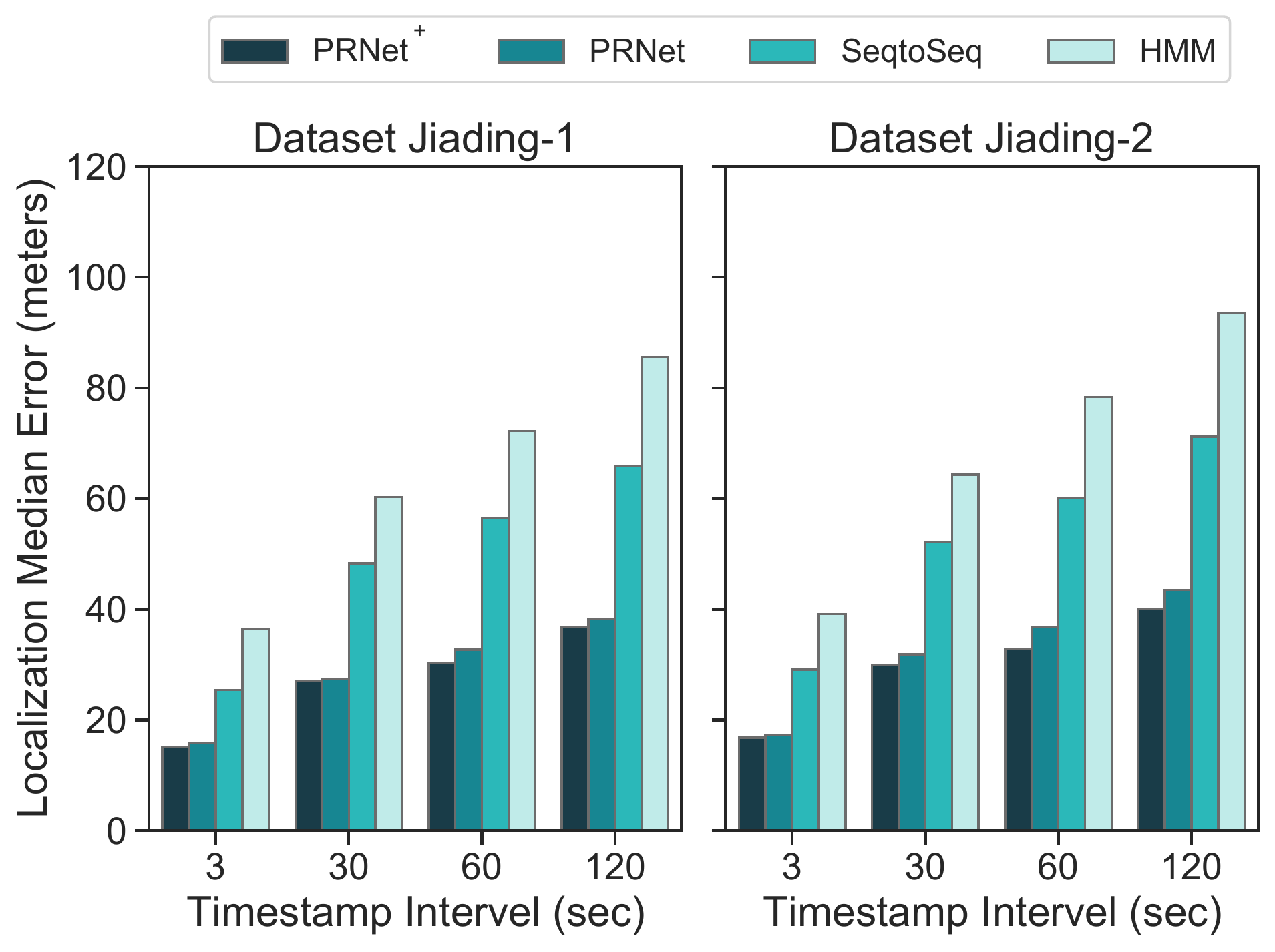}}
\end{center}
\end{minipage}
&
\begin{minipage}[t]{0.31\linewidth}
\begin{center}			
\centerline{\includegraphics[height=1.4in]{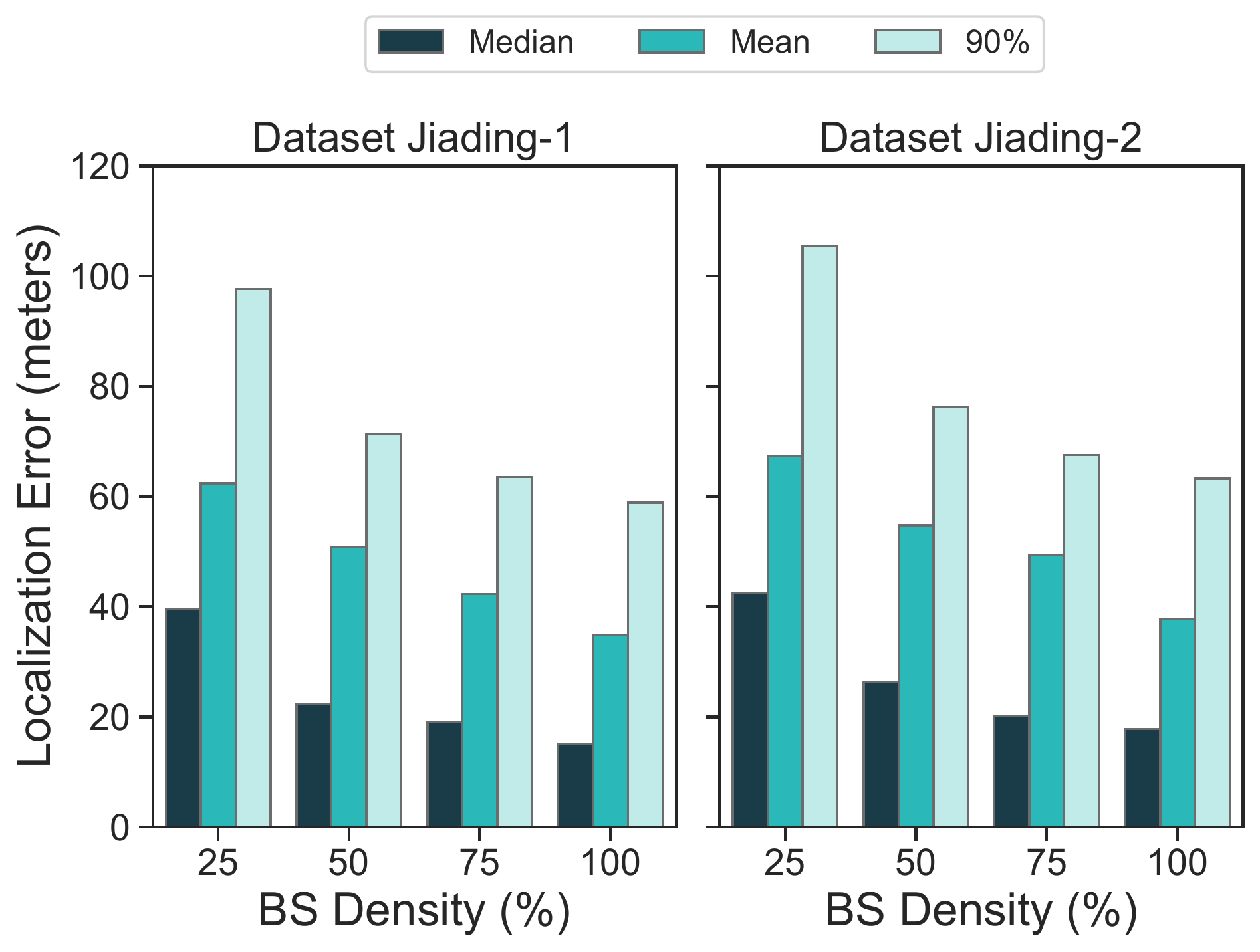}}
\end{center}
\end{minipage}
&
\begin{minipage}[t]{0.31\linewidth}
\begin{center}
\centerline{\includegraphics[height=1.4in]{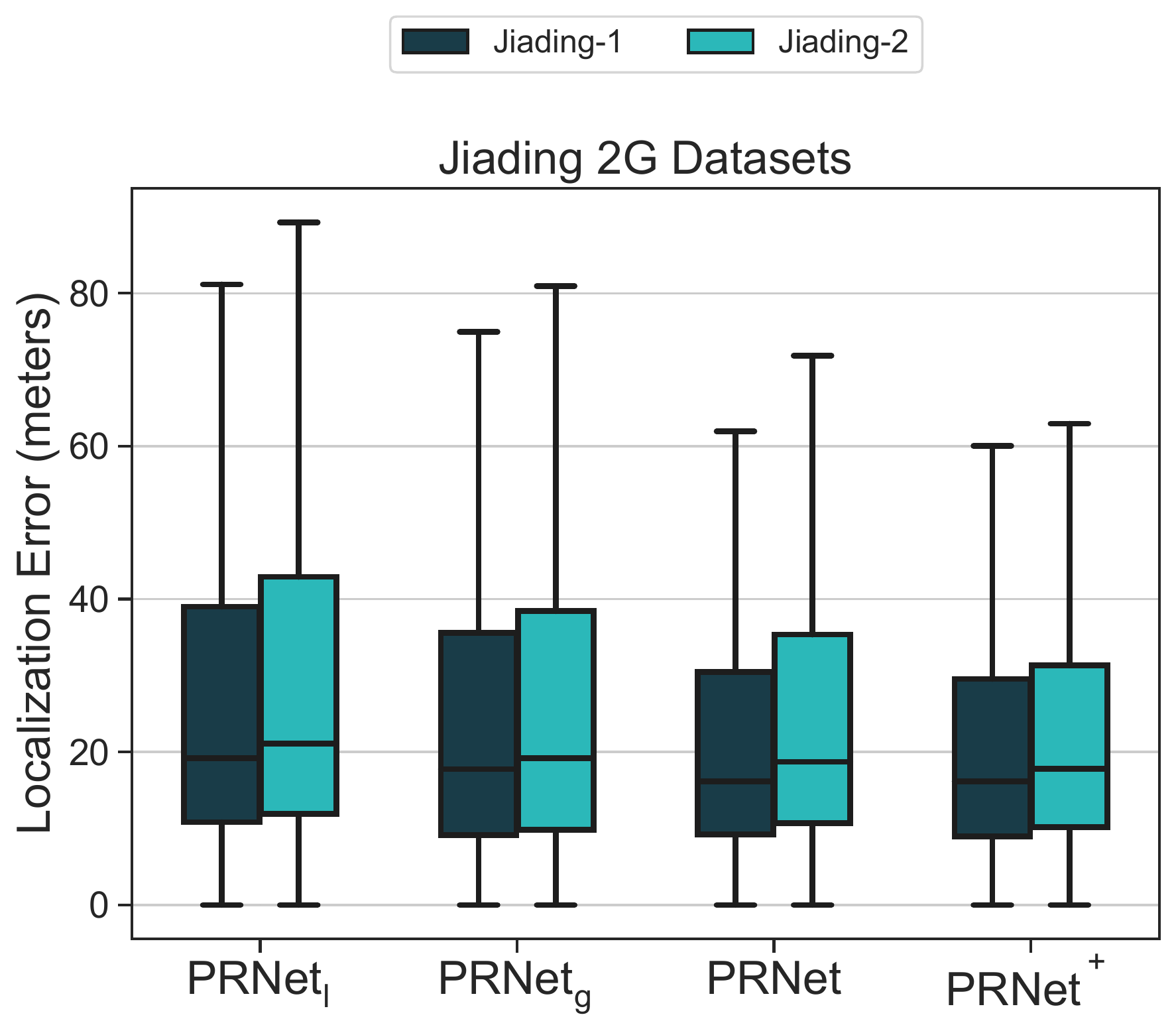}}
\end{center}
\end{minipage}
\end{tabular}
\end{center}\vspace{-5ex}
\caption{Sensitivity Study. (a)Time Interval, (b) Base Station Density, (c) Ablation Study (from left to right).}\label{exp:sense2}
\end{figure*}

\noindent \textbf{(2) \underline{Timestamp Intervals}}: In this experiment, we study the effect of timestamp intervals (used by Equation \ref{eq:aij} of time-interval attention in Section \ref{sec4.2.1}) on localization errors. From the MR samples of the \emph{Jiading-1} and \emph{Jiading-2} 2G datasets, we randomly select MR samples to make sure that the timestamp difference between neighbouring MR samples in MR sequences is smaller than a certain time interval, and we vary the time intervals from 3 seconds to 120 seconds. From Figure \ref{exp:sense2}(a), we have the following result. Firstly, a higher time interval (and thus more sparse data samples) leads to the higher prediction errors of all four sequence models. Secondly, the growth trends of median errors regarding \textsf{PRNet$^+$} and \textsf{PRNet} are rather smooth. It is mainly because \textsf{PRNet} module can capture temporal dependencies even from sparse samples. In addition, \textsf{HMM} suffers from much higher errors than \textsf{SeqtoSeq}, indicating that the deep sequence models including \textsf{SeqtoSeq}, \textsf{PRNet} and \textsf{PRNet$^+$} are better to capture long-term dependencies than the first-order HMM model. Then, even with high timestamp intervals from 30 to 120 seconds, \textsf{PRNet$^+$} and \textsf{PRNet} can still perform well, mainly due to the time interval and subsequence attention mechanism in \textsf{PRNet}/\textsf{Feature Extraction} module. 


\noindent\textbf{(3) \underline{Base Station Density}}: We now study how the density of base stations affects the prediction errors of \textsf{PRNet$^+$} by varying the percentage of base stations from 25\%-100\%. Specifically, among all base stations in a MR dataset, we randomly remove some base stations from MR samples. If a certain MR sample contains a removed base station, then the signal measurements regarding this base station are dropped from the sample. Figure \ref{exp:sense2}(b) indicates that \textsf{PRNet$^+$} leads to a competitive accuracy even if we drop 50\% base stations. Nevertheless, a lower base station density means higher errors. The reason is that sparse base stations incur spatial ambiguity of MR samples. More base stations in MR samples alternatively lead to more discriminative MR features and lower errors.

\noindent \textbf{(4) \underline{Ablation Study of {PRNet$^+$}}}: To study the importance of each component in \textsf{PRNet$^+$} for localization, we design the following variants. 1) \textsf{PRNet}$_l$ using the local predictor of \textsf{Feature Extraction Module} alone (which can be treated as a single-point-based predictor to process MR samples individually): we need to add an additional dense layer to produce the final outputs; 2) \textsf{PRNet}$_g$ using the global predictor of \textsf{Feature Extraction Module} alone: we need to reshape the  original input matrix $\bm{X}_{i,j} \in \mathbb{R}^{F \times N}$ of a MR sequence into a column vector $(F\times N)\times 1$, which is directly fed into the global predictor; 3) \textsf{PRNet} without \textsf{Transportation Mode Detection Module}. As shown in Figure \ref{exp:sense2}(c), \textsf{PRNet}$_l$ incurs a high growth of localization errors, indicating the significant contribution offered by the global predictor to capture the temporal dependencies of the entire sequence. Instead, \textsf{PRNet}$_g$ could reduce the top 90\% errors, when compared to \textsf{PRNet}$_l$. It mainly because that the dependencies among neighbouring MR samples can mitigate the prediction of outliers. Not surprisingly, \textsf{PRNet} with the local and global predictors can lead to the lowest localization errors among \textsf{PRNet} and its two variants. Since \textsf{PRNet$^+$} utilizes the detected transportation modes of MR samples by \textsf{Transportation Mode Detection} module and the shared feature representations to improve the performance of \textsf{PRNet} module, \textsf{PRNet$^+$} has the slightly lower prediction errors than \textsf{PRNet}.


\noindent\textbf{(5) \underline{Learning Rate}}: Table \ref{tab:learning_rate} shows the effect of different learning rates on \textsf{PRNet}$^+$. Learning rate determines whether the objective (loss) function can converge to a local minimum and how fast it converges. Compared to the SGD optimizer utilized by \textsf{PRNet}, Adam optimizer is quite robust for the selection of model parameters. Thus, we adopt Adam optimizer instead of SGD optimizer in \textsf{PRNet$^+$}, and vary the learning rate from 0.005 to 0.0001 under the same training epoches. The table shows that lowering the learning rate to (0.0005) makes the training more reliable. This is because lowering the learning rate makes the optimizer step towards the minimum slowly which prevents overshooting it.


\begin{table}[htb]
\scriptsize
\caption{Effect of Learning Rate}\label{tab:learning_rate}\vspace{-2ex}
\begin{tabular}{|c|l|c|c|c|c|}
\hline
\multicolumn{2}{|c|}{Learning Rate} & 0.005 & 0.001 & 0.0005 & 0.0001 \\ \hline\hline
\multirow{4}{*}{\begin{tabular}[c]{@{}c@{}}Jiading-1 \\ (2G)\end{tabular}}
 & Median Err. (meters) & 16.4 & 15.5 & 15.2 & 15.5  \\
 & Mean Err. (meters)& 38.2 & 36.5 & 34.3 & 35.1\\
 & 90\% Err. (meters)& 63.7 & 61.2 & 59.6 & 60.8 \\ \hline
\multirow{4}{*}{\begin{tabular}[c]{@{}c@{}}Jiading-2 \\ (2G)\end{tabular}} & Median Err. (meters)& 18.8 & 17.6 & 16.4 & 17.2 \\
 & Mean Err. (meters)& 38.2 & 37.8 & 37.1 & 37.9 \\
 & 90\% Err. (meters)& 66.4 & 65.8 & 64.7 & 65.2\\ \hline
\end{tabular}
\end{table}

\subsection{Visualization}
\begin{figure*}[th]
\begin{center}\vspace{-2ex}
\begin{tabular}{c c c c c c c c}
\begin{minipage}[t]{0.10\linewidth}
\begin{center}
\centerline{\subfigure[\textsf{\footnotesize{PRNet$^+$}}]{
    \label{fig:subfig:prnet} 
    \includegraphics[width=.8in]{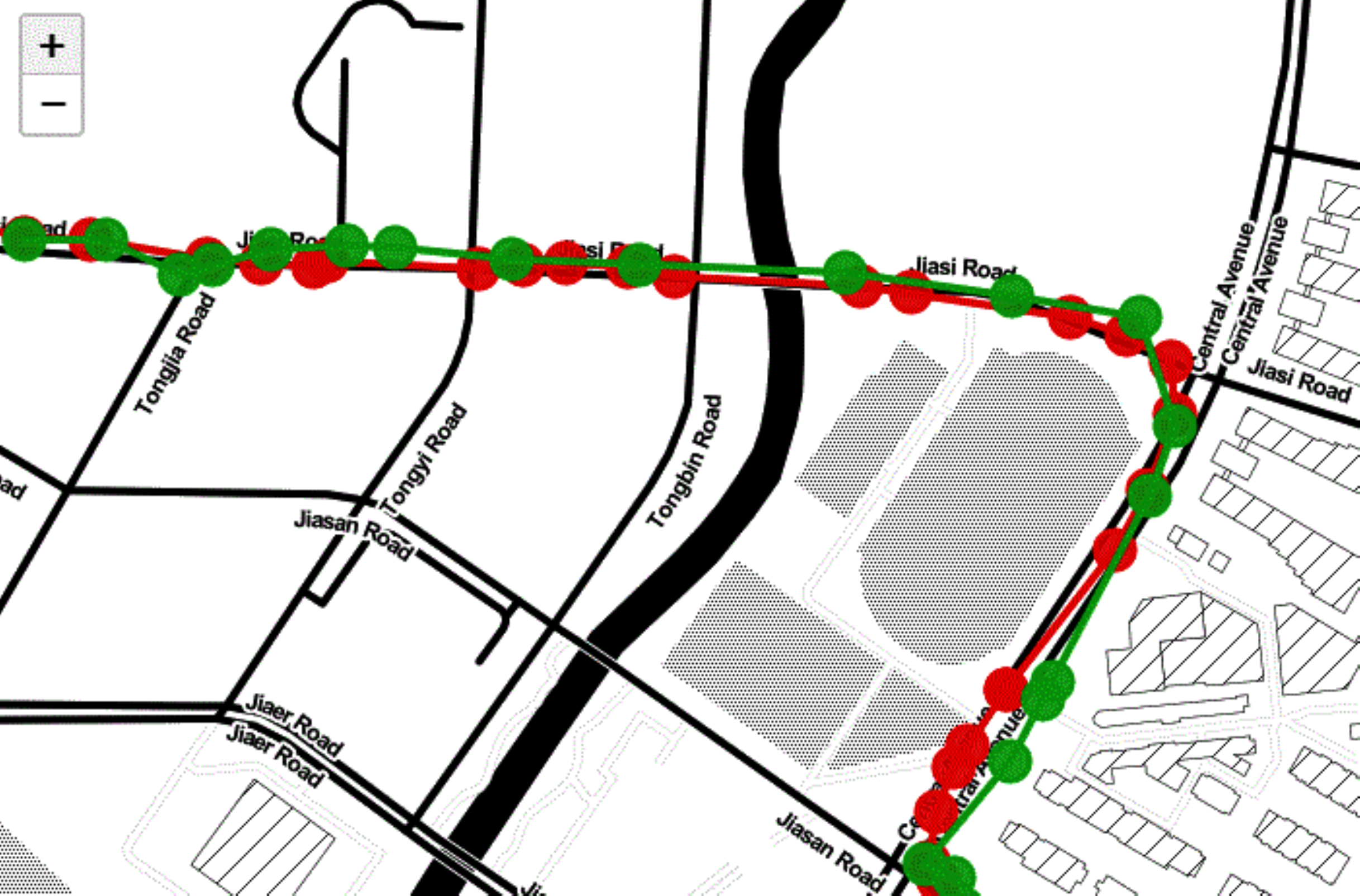}
  }}
\end{center}
\end{minipage}
&
\begin{minipage}[t]{0.10\linewidth}
\begin{center}
\centerline{\subfigure[\textsf{PRNet$^+$}\_M]{
    \label{fig:subfig:deeploc} 
    \includegraphics[width=.8in]{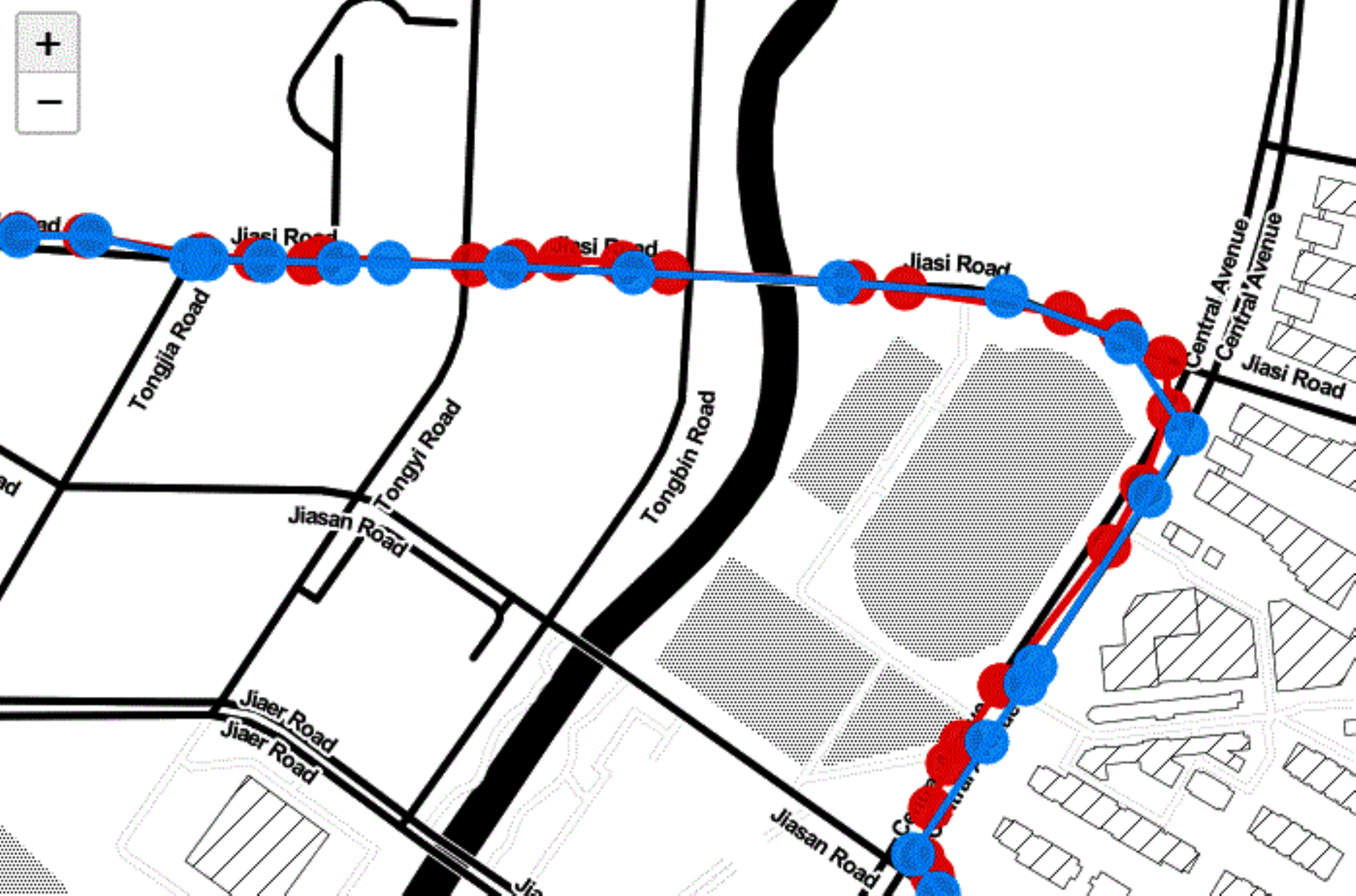}
  }}
\end{center}
\end{minipage}
&
\begin{minipage}[t]{0.10\linewidth}
\begin{center}
\centerline{\subfigure[\textsf{PRNet}]{
    \label{fig:subfig:hmm} 
    \includegraphics[width=.8in]{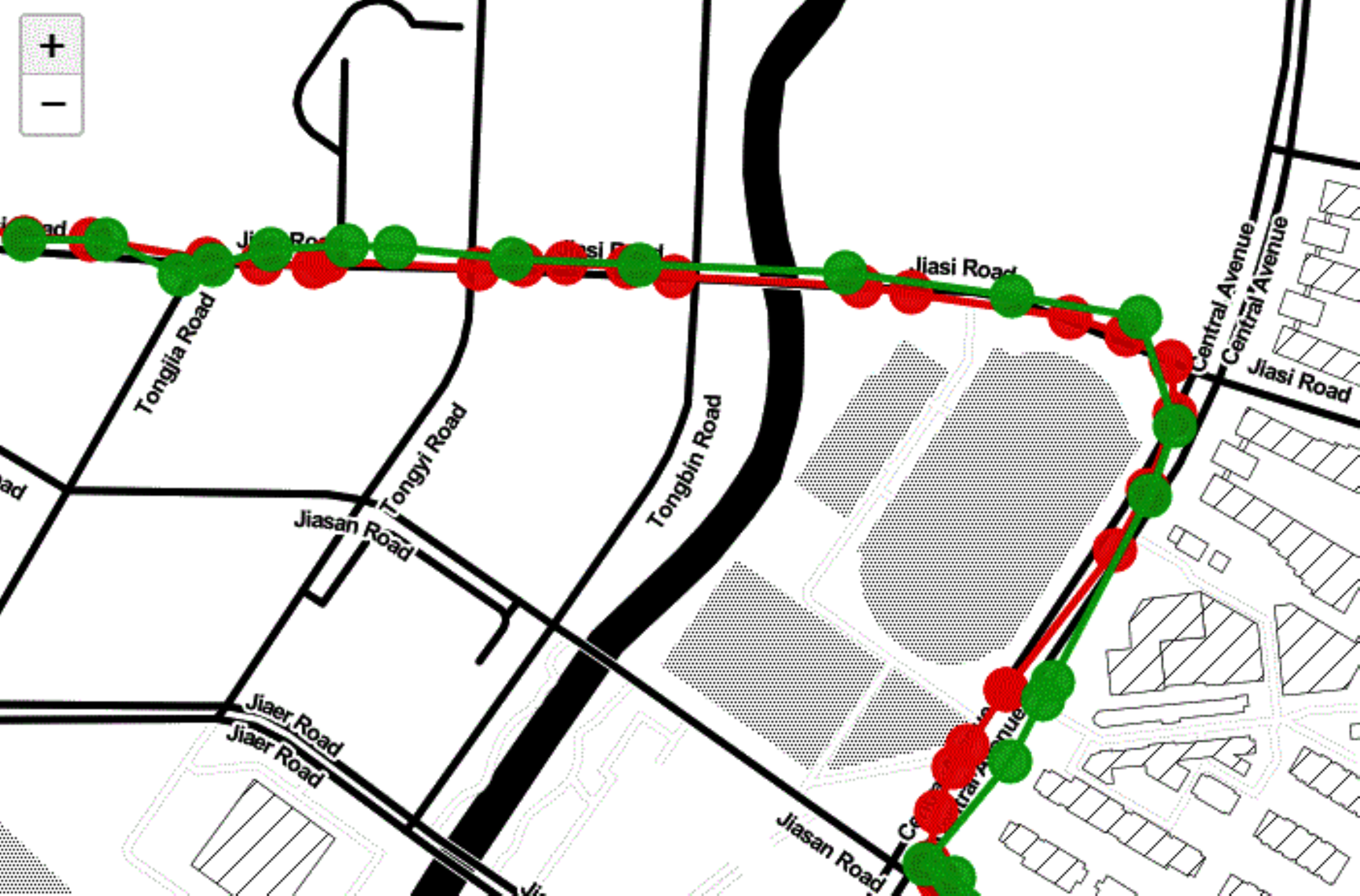}
  }}
\end{center}
\end{minipage}
&
\begin{minipage}[t]{0.10\linewidth}
\begin{center}
\centerline{\subfigure[\textsf{PRNet}\_M]{
    \label{fig:subfig:hmm} 
    \includegraphics[width=.8in]{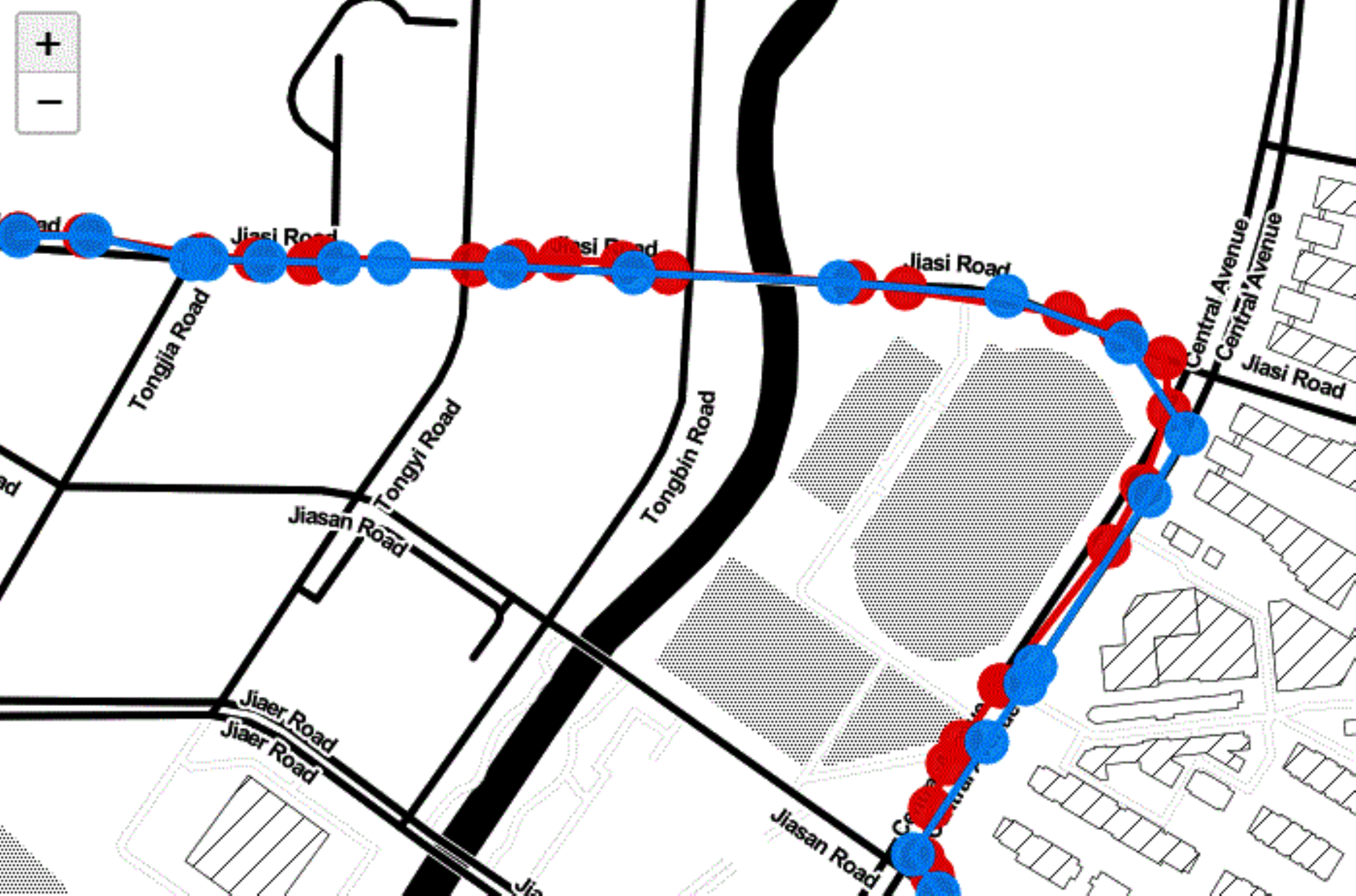}
  }}
\end{center}
\end{minipage}
&
\begin{minipage}[t]{0.10\linewidth}
\begin{center}
\centerline{\subfigure[\textsf{HMM}]{
    \label{fig:subfig:ccr} 
    \includegraphics[width=.8in]{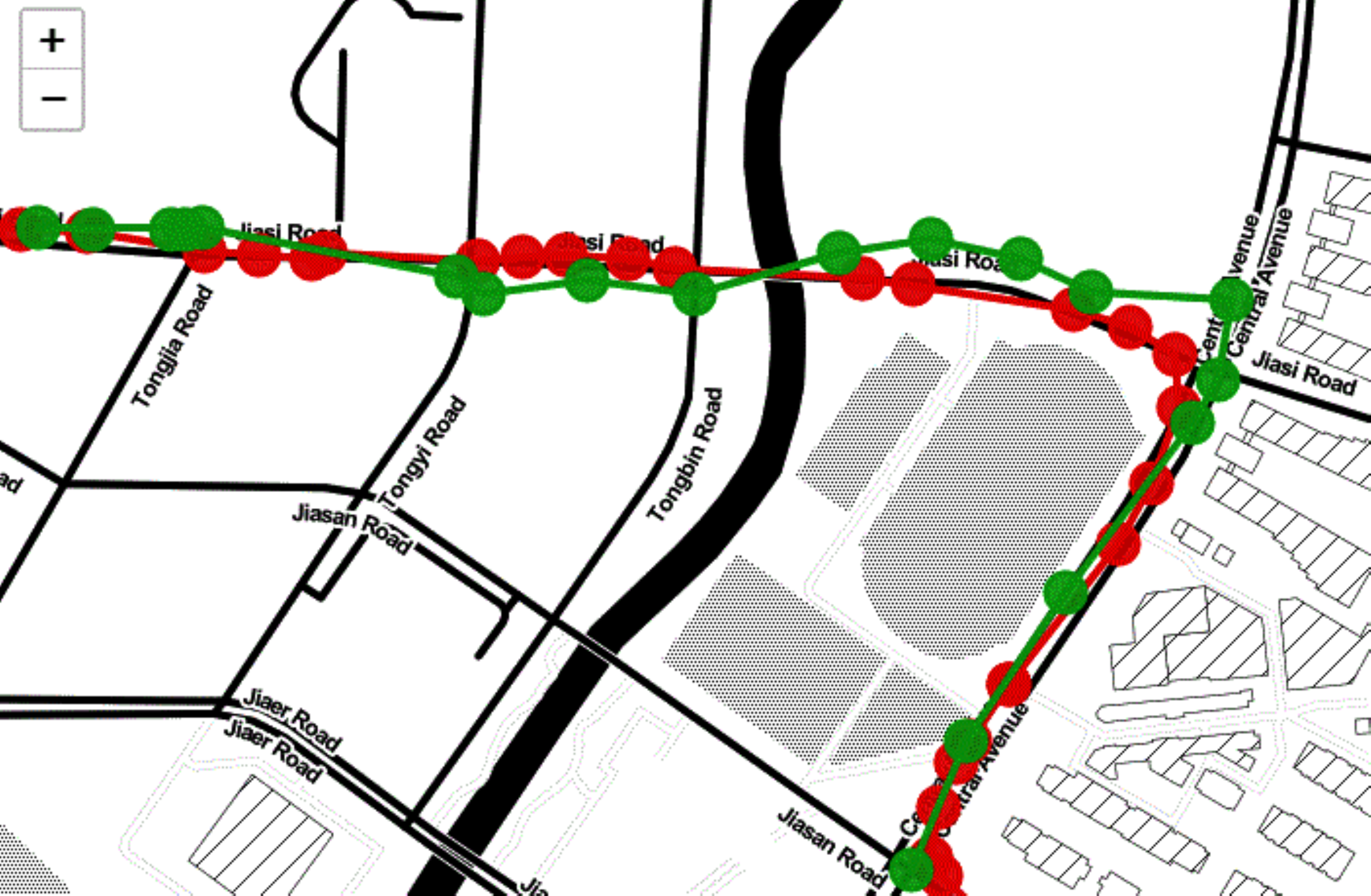}
  }}
\end{center}
\end{minipage}
&
\begin{minipage}[t]{0.10\linewidth}
\begin{center}
\centerline{\subfigure[\textsf{HMM}\_M]{
    \label{fig:subfig:ccr} 
    \includegraphics[width=.8in]{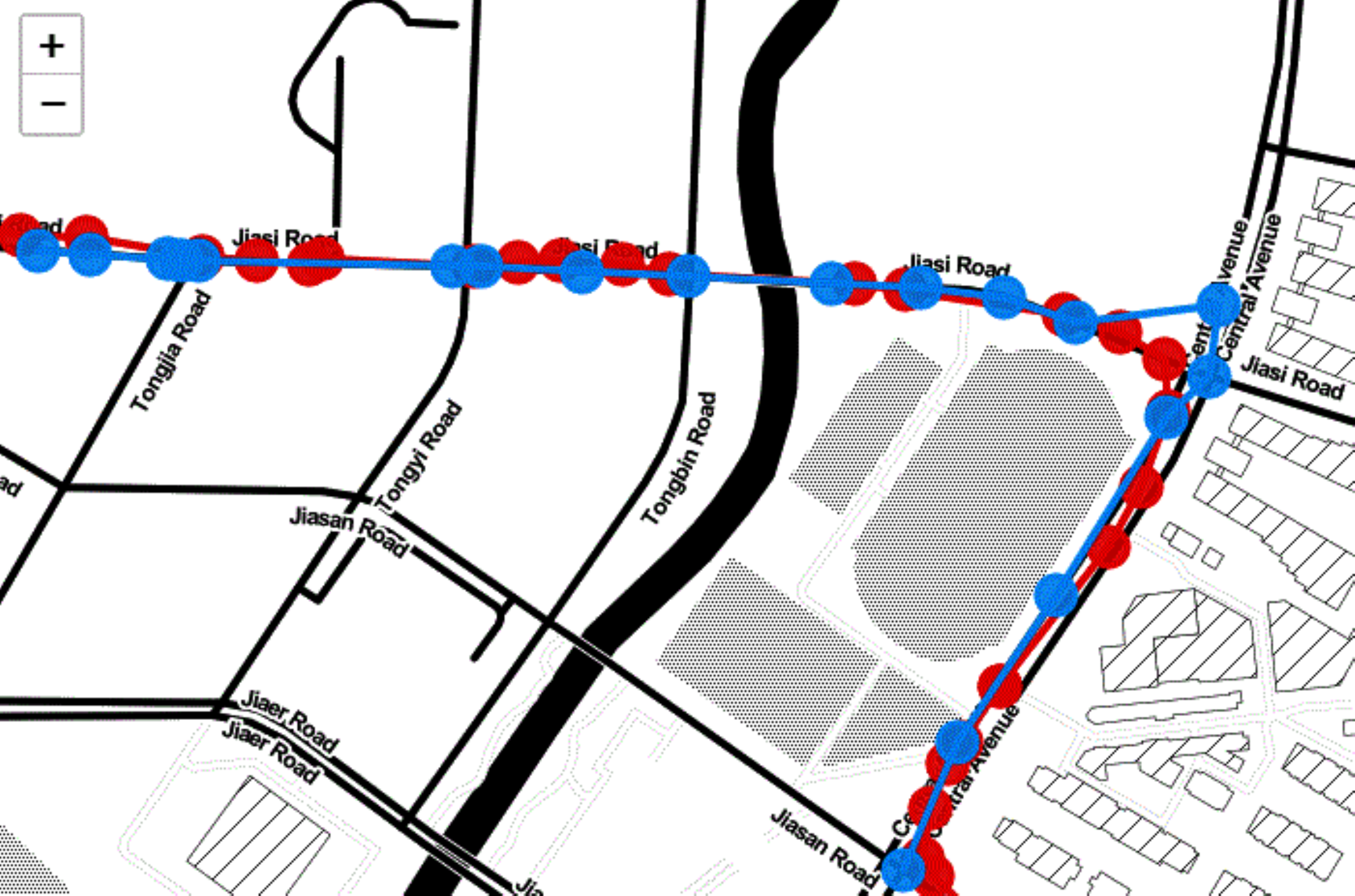}
  }}
\end{center}
\end{minipage}
&
\begin{minipage}[t]{0.10\linewidth}
\begin{center}
\centerline{\subfigure[\textsf{DeepLoc}]{
    \label{fig:subfig:nbl} 
    \includegraphics[width=.8in]{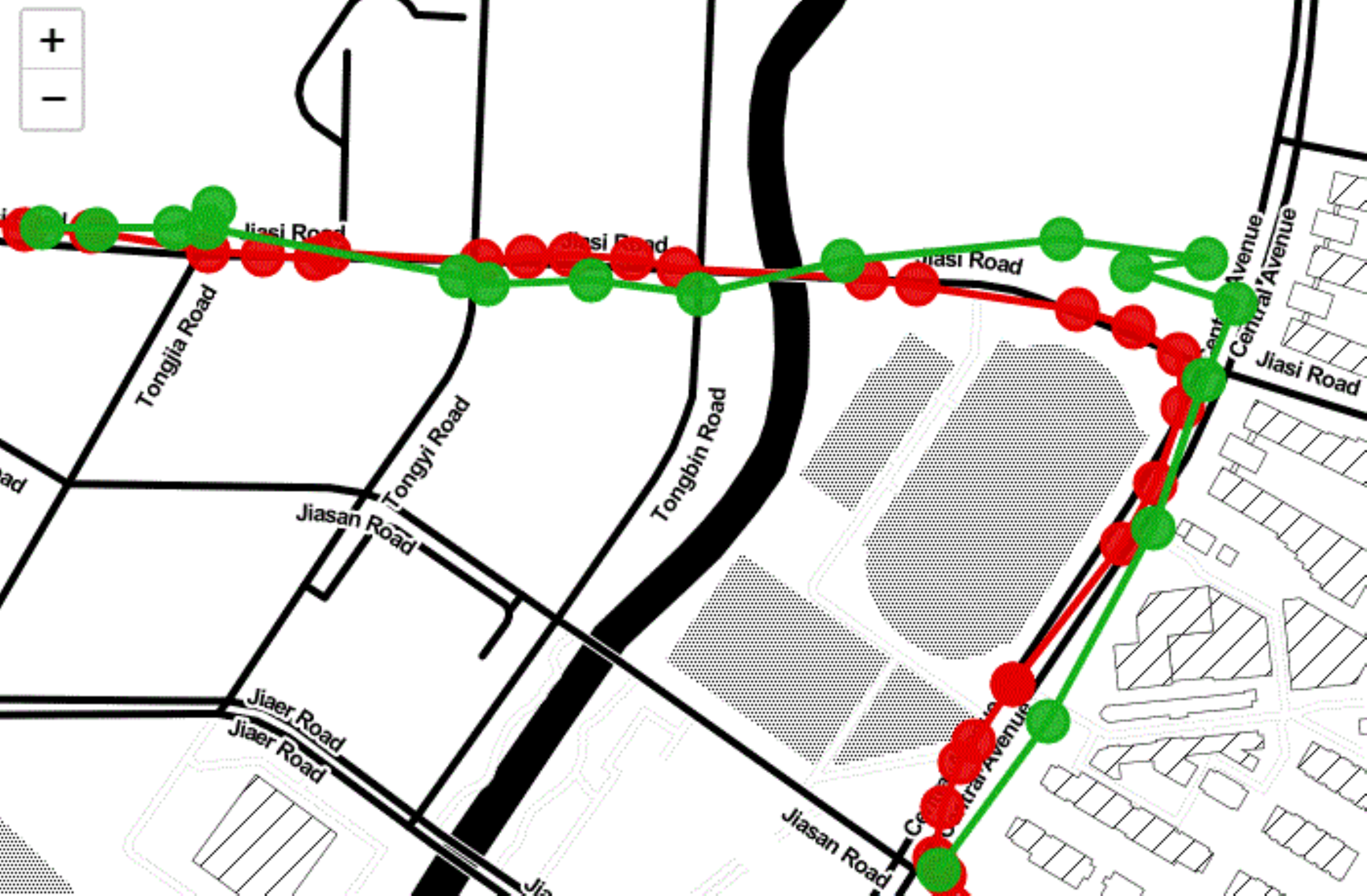}
  }}
\end{center}
\end{minipage}
&
\begin{minipage}[t]{0.10\linewidth}
\begin{center}
\centerline{\subfigure[\textsf{DeepLoc}\_M]{
    \label{fig:subfig:nbl} 
    \includegraphics[width=.8in]{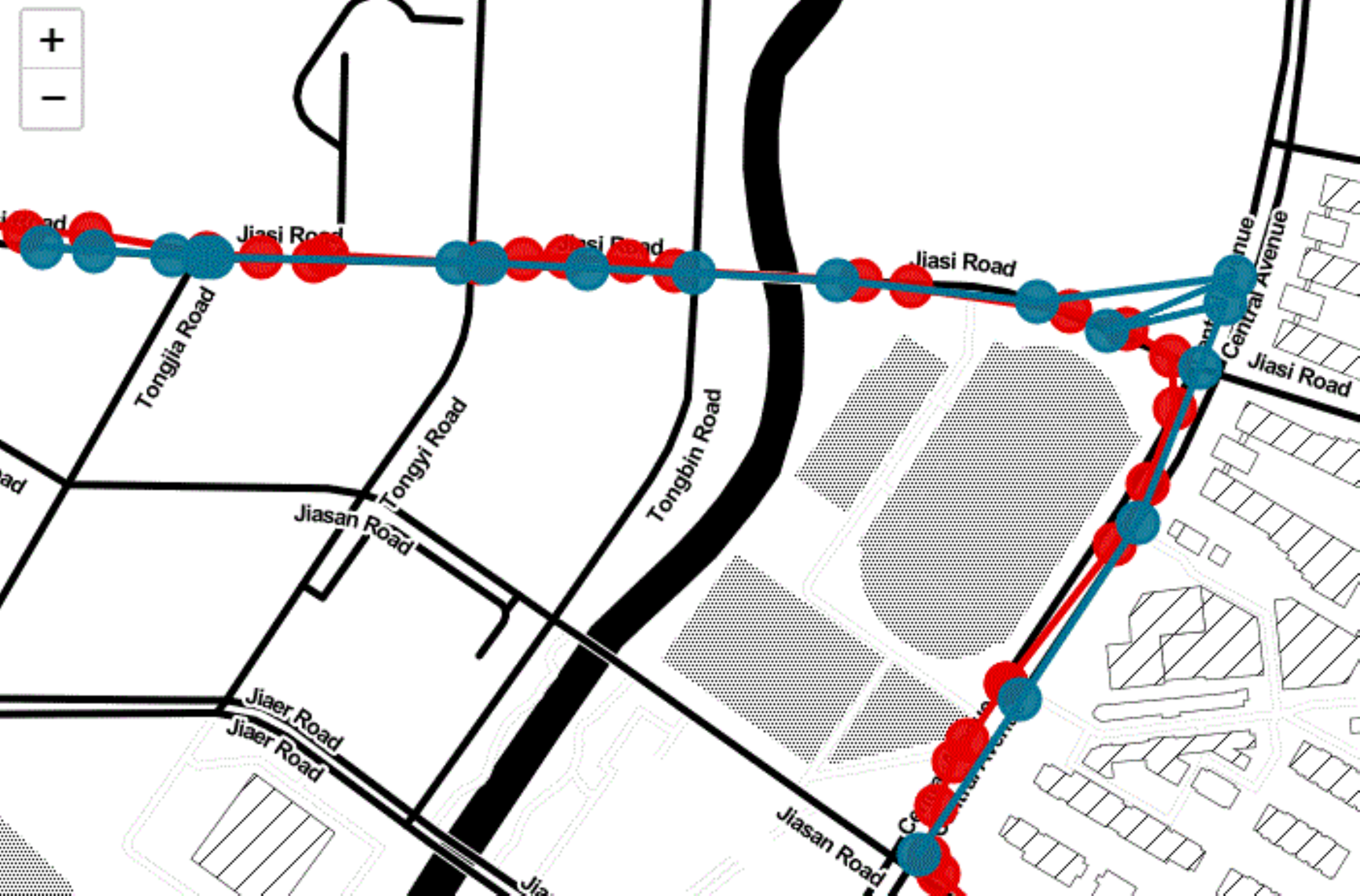}
  }}
\end{center}
\end{minipage}
\end{tabular}\vspace{-2ex}
\caption{Visualization (Map Scale 1.5:10000). Red: Ground Truth; Green: Predicted Location; Blue: Map-matching location.}\vspace{-2ex}
\label{exp:vis}
\end{center}
\end{figure*}

Finally, Figure \ref{exp:vis} visualizes the position recovery results of \textsf{PRNet$^+$} and three Telco location recovery approaches (\textsf{PRNet}, \textsf{HMM}, and \textsf{DeepLoc}) on a randomly selected small area from the \emph{Jiading-1} 2G data set. As shown in this figure, the trajectory predicted by \textsf{PRNet$^+$} is the closest to the ground truth in terms of horizontal moving directions and vertical shift out of road segments. Yet \textsf{DeepLoc} leads to the most shift in both horizontal and vertical directions. The median localization errors of \textsf{HMM}, and \textsf{DeepLoc} are very close, but \textsf{HMM} leads to a smoother trajectory of visualization results. It is mainly because \textsf{HMM} first utilized Hidden Markov Model for extracting the contextual dependencies between neighboring MR samples and next the filtering technique for post-processing.

With help of the road network, we next perform the map-matching technique \cite{ZhengCWY14a} on the predicted trajectories of the aforementioned approaches (denoted as \textsf{PRNet$^+$}\_M, \textsf{PRNet}\_M, \textsf{HMM}\_M, and \textsf{DeepLoc}\_M, respectively) and plot the map-matching results in this figure. This post-processing technique greatly improves \textsf{DeepLoc} to generate smooth trajectories. However, the map-matching technique cannot fully overcome the overall shift along the horizontal moving direction, and Figure \ref{exp:vis} has clearly visualized this issue. 

Consistent with the result in Section \ref{sec:baseline}, both \textsf{PRNet$^+$} and \textsf{PRNet} lead to acceptable results, even without this post-processing step. The results in Table 6 indicates the slight improvement of \textsf{PRNet$^+$} on \textsf{PRNet} (the 90\% error is reduced by 3.6 meters in Jiading-1 2G dataset), thus it make senses that there is no significant visualization difference between the results of \textsf{PRNet$^+$} and \textsf{PRNet}. However, \textsf{PRNet$^+$} can realize accurate fine-grained transportation mode detection for MR data and while slightly improve the performance of outdoor position recovery, which is meaningful for more urban computing scenarios.

%% file: 06-con.tex
\section{Related work}\label{sec:related}

\textbf{Trajectory Recovery}: Unlike single point-based localization, the sequence-based localization, namely trajectory recovery problem, frequently exhibit much lower localization errors. For example, AT\&T researchers \cite{RayDM16} proposed an outdoor localization approach, which first exploits the hidden markov model (HMM) for trajectory recovery and then applies particle filters to generate smooth trajectories. CCR \cite{ZhuLYZZGDRZ16} designed temporal and spatial contextual features such as moving speeds and time gap via domain expert knowledge and feature engineering expertise to greatly improve prediction accuracy. A recent work \cite{8939387}, though achieving rather precisely localization result, requires massive third-party GPS trajectories as the prior to precisely estimate the HMM emission probability. In addition, the work \cite{DBLP:journals/paapp/TranZG16} proposed a regularization framework to reconstruct mobile trajectories from sparse location fingerprints by enforcing two properties including spatial and temporal smoothness. Similar to \cite{DBLP:journals/paapp/TranZG16}, map-matching techniques and their variants \cite{ZhangRDZXA18,HuangRZZYZ18}
attempt to recover an entire trajectory from a set of (sparse) locations, e.g., GPS coordinates. These works, with help of road network constraints, project the locations onto road networks to correct outlier locations (e.g., noisy GPS points) and recover an entire trajectory. Unlike \cite{DBLP:journals/paapp/TranZG16,DBLP:journals/paapp/TranZG16}, our work does not require the availability of either sparse location points within input trajectories or road networks.

\textbf{Deep Learning for Localization Systems}:
Recently, deep neural networks have been used in indoor or outdoor localization systems. Firstly, in dynamic indoor environment, the Deep Belief Network (DBN) and Gaussian-Bernoulli Restricted Boltzmann Machines have been utilized in fingerprinting-based indoor localization to increase estimation accuracy and reduce generalization error \cite{DBLP:conf/icufn/FelixSN16}. In addition, to process the RSS time-series acquired from wireless local area network (WLAN) access points, the previous work \cite{DBLP:conf/iscc/IbrahimTE18} leveraged a convolutional neural network (CNN) which is fed with RSS feature matrices for indoor localization by extracting the temporal dependencies between the last RSS readings. DeepFi \cite{DBLP:journals/tvt/WangGMP17} is another deep-learning-based indoor fingerprinting system based on channel state information (CSI) rather than RSS for indoor localization. DeepFi explores the features of wireless channel data and obtains the optimal weights of DNN as fingerprints which effectively describe the characteristics of CSI for each location and help reduce noise. These fingerprint-based approaches typically require sufficient samples to construct fingerprint database, and do not work well for our problem with heterogenous samples. Finally, for outdoor Telco localization, DeepLoc \cite{DBLP:conf/gis/ShokryTY18} employs a data augmentation technique to tackle the challenges of noisy data and insufficient training samples. In summary, all of these DNN-based works typically perform single-point-based localization though they improve localization result via various contextual spatio-temporal features \cite{DBLP:conf/iscc/IbrahimTE18} and instead we focus on a sequence-based DNN model.

\textbf{Transportation Mode Detection}: Many literature works have studied the problem of transportation mode detection on GPS trajectory data. For example, the previous works \cite{DBLP:conf/gis/StennethWYX11, DBLP:journals/tweb/ZhengCLXM10} first extract the features such as velocity and then train a machine learning-based detection model. Some works employ deep learning techniques for transportation mode detection, e.g., the CNN-based and LSTM-based detection \cite{DBLP:journals/corr/abs-1804-02386,DBLP:conf/ijcai/SongKS16}. Instead of GPS data, some works perform transportation mode detection via the readings from accelerators and gyroscopes \cite{DBLP:journals/imwut/ChenSHN17,DBLP:conf/sensys/HemminkiNT13}.

Similar to our work, Monitor \cite{DBLP:conf/vtc/Al-HusseinyY12}, and MonoSense \cite{DBLP:conf/vtc/AbdelAzizY15} detect transportation modes on the data generated by cellular networks with help of base station IDs and the associated received signal strength (RSSI) to extract statistical features such as velocity. Nevertheless, our work exploits learned features rather than statistical features for better accuracy. Some works \cite{DBLP:conf/huc/SohnVLCCSCHGL06,DBLP:journals/tosn/ReddyMBEHS10} incorporate cellular data and other sensor data collected from mobile devices to detect transportation mode. CellTrans \cite{DBLP:journals/imwut/Zhao0LZ019} infers that mobile users take either public transportation tools or private car at urban scale by extracting meaningful mobility features from cellular data. Such features are computed by the positions of connected base stations from Mobile flow records (MFRs) at a coarse-grained level (each trip of a user is tagged with a certain mode). Instead, we perform fine-grained transportation mode detection with one mode per MR sample.


\section{Conclusion}\label{sec:con}
In this paper, we proposed a multi-task learning framework called \textsf{PRNet}$^+$. \textsf{PRNet}$^+$ first ensembles the power of \textsf{CNN}, \textsf{LSTM}, and two attention mechanisms to properly leverage all of the local, short- and long-term spatial and temporal dependencies to learn MR features. On the learned MR features, \textsf{PRNet}$^+$ performs two learning tasks (outdoor location recovery and transportation mode detections) with help of a joint loss function. Our extensive evaluation shows that \textsf{PRNet}$^+$ greatly outperforms state-of-the-art localization approaches and alternative variants of \textsf{PRNet}$^+$ on the datasets collected in three representative areas in Shanghai. The promising result of \textsf{PRNet}$^+$ inspires future studies on issues such as how to use the recovered trajectories from Telco MR data for human mobility analysis and the potentials of applying \textsf{PRNet}$^+$ to indoor localization and forthcoming 5G networks.

\textbf{Acknowledge}: This work is partially supported by National Natural Science Foundation of China (Grant No. 61972286, No. 61772371).